\documentclass[iop]{emulateapj}
\usepackage{bm}
\usepackage{threeparttable}
\usepackage{morefloats}
\usepackage{CJK}
\def\bfnabla{{\mbox{\boldmath $\nabla$}}}

\def\msun{M_\odot}
\def\mbh{M_{\rm{BH}}}
\def\Medd{\dot{M}_{\rm{Edd}}}
\def\Ledd{L_{\rm{Edd}}}

\renewcommand\bv{{\mbox{\boldmath $v$}}}
\newcommand\bb{{\mbox{\boldmath $B$}}}
\newcommand\bP{{\mbox{\boldmath $P$}}}
\newcommand\bn{{\mbox{\boldmath $n$}}}

\newcommand\br{{\mbox{\boldmath $r$}}}
\newcommand\bF{{\mbox{\boldmath $F$}}}

\def\<{\,\langle\langle}
\def\>{\,\rangle\rangle}

\begin{document}
\begin{CJK*}{UTF8}{gbsn}

\shortauthors{Y.-F. Jiang et al.}
\author{Yan-Fei Jiang(姜燕飞)\altaffilmark{1}, James M. Stone\altaffilmark{2} \& Shane W. Davis\altaffilmark{3}}
%\author{Yan-Fei Jiang\altaffilmark{1,2}\footnote{Einstein Fellow}, James M. Stone\altaffilmark{1} \& Shane W. Davis\altaffilmark{3}}
\affil{$^1$kavli institute for theoretical physics, Kohn Hall, University of California, Santa Barbara 93106, USA} 
\affil{$^2$Department of Astrophysical Sciences, Princeton
University, Princeton, NJ 08544, USA} 
\affil{$^3$Department of Astronomy, University of Virginia, P.O. Box 400325, Charlottesville, VA 22904-4325, USA}

\title{Super-Eddington Accretion Disks around Supermassive black Holes}

\begin{abstract}
	
We use global three dimensional radiation magneto-hydrodynamical simulations 
to study accretion disks onto a $5\times 10^8\msun$ black hole with accretion rates 
varying from $\sim 250\Ledd/c^2$ to $1500\Ledd/c^2$. We form the disks with torus centered at $50-80$ 
gravitational radii with self-consistent turbulence initially generated 
by the magneto-rotational instability. We study cases with and without 
net vertical magnetic flux. 
The inner regions of all disks have radiation 
pressure $\sim 10^4-10^6$ times the gas pressure. Non-axisymmetric density waves that steepen into spiral shocks 
form as gas flows towards the black hole. In simulations without net vertical magnetic flux, Reynolds stress generated by the spiral shocks 
are the dominant mechanism to transfer angular momentum.  
Maxwell stress from MRI turbulence can be larger than the Reynolds stress only when net vertical 
magnetic flux is sufficiently large. Outflows are formed with speed $\sim 0.1-0.4c$. 
When the accretion rate is smaller than $\sim 500\Ledd/c^2$, outflows start around $10$ gravitational radii and the radiative efficiency 
is $\sim 5\%-7\%$ with both magnetic field configurations. With accretion rate reaching $1500\Ledd/c^2$, 
most of the funnel region close to the rotation axis becomes optically thick and the outflow 
only develops beyond $50$ gravitational radii. The radiative efficiency is reduced to $1\%$.  
We always find the kinetic energy luminosity associated with the outflow is only 
$\sim 15\%-30\%$ of the radiative luminosity. The mass flux lost in the outflow is $\sim 15\%-50\%$ of the net mass accretion rates. 
We discuss implications of our simulation results on the observational properties of these disks.

\end{abstract}


\maketitle

\section{Introduction}
Active Galactic Nuclei (AGNs)  are believed to be powered by accretion of gas onto 
the central supermassive black hole (SMBH) with black hole mass $\mbh\sim 10^6-10^{10}\msun$. Although the observed bolometric luminosity for most AGNs are estimated to be smaller than the Eddington value $\Ledd=4\pi G\mbh c/\kappa_{\rm es}$ for electron scattering opacity $\kappa_{\rm es}$ \citep[e.g.][]{Kollmeier2006}, super-Eddington accretion is expected in many circumstances. For example, narrow-line seyfert 1 galaxies (NLS1) are widely believed to host supermassive black holes with accretion rate close or above the Eddington limit \citep[][]{Poundsetal1995,Komossaetal2006,Jinetal2016,Jinetal2017}. The weak emission-line quasars have also been suggested to host super-Eddington accretion disks \citep[][]{Luoetal2015}. The existence of $\sim 10^9\msun$ black holes around redshift $z\sim 7$ \citep[][]{Mortlocketal2011} implies that super-Eddington accretion around supermassive black holes must happen for at least a fraction of the time in order for these black holes to gain so much mass before this redshift for any reasonable model of black hole seeds \citep[][]{Volonterietal2015,BegelmanVolonteri2017}. Tidal disruption events are also thought to trigger super-Eddington accretions onto supermassive black holes if the stripped mass from the stars can circularize and form an accretion disk efficiently \citep[]{Rees1988}.  
Because feedback from AGNs is an important process to regulate the evolution and structures of the host galaxies \citep[e.g.,][]{CiottiOstriker2007,Ciottietal2010,KormendyHo2013}, it is important to understand when and how AGNs can accrete above the Eddington limit. Although this process may only happen for a small fraction of the duty cycles for AGNs, it can produce significant amount of radiation and outflow from the disks, which may have important feedback effects.

In order to tell whether the disks in AGNs are accreting above the Eddington limit or not, it is useful to predict what to expect from this kind of   disks, as direct measurements of black hole mass and bolometric luminosity for AGNs are usually very uncertain. The slim disk model  
 \citep[][]{Abramowiczetal1980} is often used to describe accretion disks in this regime and there have been a number of numerical simulations 
 to study the properties of super-Eddington accretion disks recently \citep[][]{Ohsugaetal2005,Jiangetal2014c,McKinneyetal2014,Sadowskietal2014}. However, most of these simulations are designed for stellar mass black holes with electron scattering assumed to be the dominant opacity with a few exceptions \citep[][]{Turner2004,Jiangetal2013b,Jiangetal2016a}. The general hope is that 
 these disks in AGNs are just scaled up versions of disks in X-ray binaries and some of the numerical simulation results are extrapolated to AGN even when they are run with stellar mass black hole parameters. However, although the inner regions of super-Eddington disks for both X-ray binaries and AGNs are radiation pressure dominated, the ratio between radiation pressure and gas pressure is normally smaller than $\sim 100$ in X-ray binaries while this ratio in AGNs can easily reach $\sim 10^4-10^5$ \citep[][]{ShakuraSunyaev1973}. The significantly enhanced radiation pressure might affect the properties of disks in many interesting ways that are unique to AGNs. For example, when the ratio between radiation pressure and gas pressure is larger than the ratio between proton mass and electron mass, bulk comptonization can work efficiently to change the spectra of AGN disks and may explain the soft X-ray access in AGNs \citep[][]{Socratesetal2004,Kaufmanetal2017}. Strongly radiation pressure dominated flow is also well known 
to be highly compressible \cite[][]{Turneretal2003,Jiangetal2013b} and radiative damping can significantly change the properties of turbulence generated by 
magneto-rotational instability (MRI, \citealt{BalbusHawley1991}). Particularly, the ratio between Maxwell stress and total thermal pressure in the strongly radiation pressure dominated compressible flow is usually smaller than the value found by MRI simulations without strong radiation pressure \citep[][]{Jiangetal2013b}. Finally, the temperature and density regime appropriate for AGN disks, the iron opacity bump changes the structures and stability of AGN disks significantly \citep[][]{Jiangetal2016a}. These new physical phenomena in the strongly radiation pressure dominated flow suggest that in order to understand the properties of AGN disks, we have to carry out the simulations with realistic parameters to achieve this large ratio between radiation pressure and gas pressure with realistic opacities. The focus of this paper is to show a set of numerical simulations to achieve this goal for the first time. 

MRI turbulence is now widely believed to be responsible for the angular momentum transfer in the fully ionized AGN disks. 
All the previous numerical simulations of AGN disks with self-consistent thermal properties \citep[][]{Turner2004,Jiangetal2016a} are done for local patches of disks under the shearing box approximation to study properties of the saturation state of MRI. Global simulations are necessary to study the radial structures of the disks and outflow, particularly in the super-Eddington regime. Global simulations also allow for other hydrodynamic mechanisms of angular momentum transfer to manifest \citep[][]{FromangLesur2017}, which may not be correctly captured by the local shearing box simulations. One important result of this paper is to show that global coherent spiral density waves can be excited in AGN disks, which may play an important role for angular momentum transfer when the saturation state of MRI turbulence is suppressed by radiative damping. Global simulations are significantly more expensive compared with local shearing box simulations if we want to resolve the accretion disk region while capture the large scale outflow. To achieve this goal, we use the static mesh refinement technique in the recently developed radiation MHD code {\sf Athena++} (Stone et al., in preparation). For the simulations in this paper, we place the initial torus as far as we can from the black hole (typically 80 gravitational radii) and let the disk form from vacuum self-consistently.

Properties of MRI turbulence are well known to depend on the configurations of the initial seed magnetic fields since the first 3D simulation of MRI turbulence done by \cite{HGB1995}. When there are net poloidal magnetic fields through the disk, both local and shearing box simulations show that the ratio between Maxwell stress and thermal pressure is larger compared to the case without net vertical magnetic fields \citep[][]{BaiStone2013,SuzukiInutsuka2014,Salvesenetal2016,ZhuStone2017}. Particularly, efficiency of radiative damping on MRI turbulence \citep[][]{Jiangetal2013b} may also depend on the amount of poloidal magnetic fields. Our previous simulation for super-Eddington accretion disk of stellar mass black holes \citep[][]{Jiangetal2014c} found that magnetic buoyancy caused by MRI turbulence could enhance the advective cooling along the vertical direction of the disk, which increased the radiative efficiency of super-Eddington accretion disks compared to the classical slim disk model. It is also interesting to see how different magnetic field configurations change the importance of this effect. 
% The amount of magnetic buoyancy that can happen in the disk depends on the ratio between magnetic pressure and thermal %pressure, which again increases with larger poloidal magnetic fields. 
%As magnetic field configurations in AGN disks are generally unknown, we will show simulations with and without net poloidal magnetic fields for %different mass accretion rates. 

Because it is very expensive to solve radiative transfer equation directly, flux limited diffusion (FLD) or M1 closure scheme are usually adopted in most of previous numerical simulations of super-Eddington accretion disks \citep[][]{Ohsugaetal2005,Sadowskietal2014,McKinneyetal2014}. It is usually argued that these approximate numerical schemes should work fine near the optically thick part of the disk, where diffusion approximation should be valid. However, as pointed out by \cite{Kaufmanetal2017}, these approximate local closure schemes does not include the second 
order energy exchange in terms of the ratio between flow velocity and the speed of light, which is essential for radiation viscosity. This effect may be safely neglected when the ratio between radiation and gas pressure is not too large, which is not true for AGN disks. Instead of using these approximate closure scheme, we solve the time dependent radiative transfer equation directly based on the numerical algorithm developed by 
\cite{Jiangetal2014c}. This allows us to capture the physics of radiation viscosity correctly in this strongly radiation pressure dominated flow for a wide range of optical depth.  

The paper is organized as follows. We list the equations we solve in Section \ref{sec:eqn} and describe the simulation setup in Section \ref{sec:setup}. The main properties of the disks from the simulations are given in Section \ref{sec:result}. 
Implications of the simulations are discussed in Section \ref{sec:discussion}. Finally, the results are summarized in 
Section \ref{sec:summary}.

\section{Equation}
\label{sec:eqn}
We solve the following set of ideal MHD equations coupled with the time dependent radiative transfer equations as 
\begin{eqnarray}
\frac{\partial\rho}{\partial t}+\bfnabla\cdot(\rho \bv)&=&0, \nonumber \\
\frac{\partial( \rho\bv)}{\partial t}+\bfnabla\cdot({\rho \bv\bv-\bb\bb+{{\sf P}^{\ast}}}) &=&-\bm{ S_r}(\bP)-\rho\bfnabla\phi,\  \nonumber \\
\frac{\partial{E}}{\partial t}+\bfnabla\cdot\left[(E+P^{\ast})\bv-\bb(\bb\cdot\bv)\right]&=&-cS_r(E)-\rho\bv\cdot\bfnabla\phi,  \nonumber \\
\frac{\partial\bb}{\partial t}-\bfnabla\times(\bv\times\bb)&=&0.
\label{MHDEquation}
\end{eqnarray}
 \begin{eqnarray}
 \frac{\partial I}{\partial t}+c\bn\cdot\bfnabla I&=&S(I,\bn).
  \label{RTequation}
 \end{eqnarray}
Here, $\rho,\ \bb, \bv$ are density, magnetic field and flow velocity, ${\sf P}^{\ast}\equiv(P_g+B^2/2){\sf I}$ (with ${\sf I}$
the unit tensor), $P_g$ is the gas pressure, and the magnetic permeability $\mu=1$.  The total gas energy density is
\begin{eqnarray}
E=E_g+\frac{1}{2}\rho v^2+\frac{B^2}{2},
\end{eqnarray}
where $E_g=P_g/(\gamma-1)$ is the internal gas energy density with adiabatic index $\gamma=5/3$. 
Gas temperature is calculated as $T=P_g/R_{\text{ideal}}\rho$, where
$R_{\text{ideal}}$ is the ideal gas constant with mean molecular weight $\mu=0.6$.
 
These equations are similar as the ones we solve in \cite{Jiangetal2014c,Jiangetal2014b}, except that 
we do not expand the radiation source terms $\bm{ S_r}(\bP), S_r(E)$ and  $S(I,\bn)$ to the first order of $v/c$. 
Instead, the lab frame specific intensity $I(\bn)$ with angle $\bn$
is first transformed to the co-moving frame intensity $I_0(\bn_0)$ with angle $\bn_0$ via Lorentz transformation 
\citep[e.g.,][]{MihalasMihalas1984}. 
The source terms to describe the interactions between gas and radiation in the co-moving frame are 
\begin{eqnarray}
 S(I_0,\bn_0)&=&c\rho\kappa_{aR}\left(\frac{a_rT^4}{4\pi}-I_0\right) +c\rho\kappa_s\left(J_0-I_0\right)\nonumber\\
 &+&c\rho(\kappa_{aP}-\kappa_{aR})\left(\frac{a_rT^4}{4\pi}-J_0\right) \nonumber\\
 &+&c\rho\kappa_{s}\frac{4(T-T_r)}{T_e}J_0.
 \label{eqn:radsource}
\end{eqnarray}
Here $\kappa_{aR}$ and $\kappa_{aP}$ are the Rosseland and Planck mean absorption opacities while 
$\kappa_s$ is the electron scattering opacity. The angular quadrature of the specific intensity in the 
co-moving frame is $J_0=\int I_0(\bn_0)d\Omega_0$. The last term in the above equation 
is an approximate way to mimic the energy exchange via Compton scattering 
\citep[][]{Hiroseetal2009,Jiangetal2013c,Jiangetal2014c}, 
where $T_r\equiv (4\pi J_0/a_r)^{1/4}$ with the radiation constant $a_r=7.57\times 10^{15}$ erg cm$^{-3}$ K$^{-4}$ 
is the radiation temperature and $T_e=5.94\times 10^{9}$ K is the effective temperature for electron rest mass. 
All the source terms are added implicitly as in \cite{Jiangetal2014b}. After the specific intensities $I_0(\bn_0)$ are updated 
in the co-moving frame, they are transformed back to the lab frame via Lorentz transformation. Then the 
radiation momentum and energy source terms $\bm{ S_r}(\bP)$ and $S_r(E)$ are calculated by the differences between 
the angular quadratures of $I(\bn)$ in the lab frame before and after adding the source terms. 
It can be easily confirmed that for the absorption 
source terms in equation \ref{eqn:radsource}, 
only the Planck mean absorption opacity $\kappa_{aP}$ enters the energy equation while the momentum 
source term only contains the Rosseland mean opacity $\kappa_{aR}$, which are consistent with the 
radiation moment equations we solve in \cite{Jiangetal2012} and \cite{Jiangetal2013c} to the first order 
$v/c$. The source term for Compton scattering also only affects the radiation energy density but not 
momentum in the co-moving frame. The transport step in equation (\ref{RTequation}) is solved  
in the same way as in \cite{Jiangetal2014b}. We are able to reproduce all the test problems described 
in \cite{Jiangetal2014b} and the new algorithm has been successfully used to study stream-stream collisions 
in tidal disruption events \citep[][]{Jiangetal2016b}.

Following \cite{Jiangetal2016a}, in order to capture the important effects caused by the iron opacity bump 
in AGN accretion disks, we calculate the total Rosseland mean opacity based on the same opacity table 
as in Figure 2 of \cite{Jiangetal2015}. We then subtract the electron scattering value 
$\kappa_{\rm es}=0.34$ cm$^2$ g$^{-1}$ 
to get the Rosseland mean absorption opacity $\kappa_{aR}$. As we do not have any Planck mean 
opacity table, we only include the Planck mean free-free absorption opacity 
$\kappa_{aP}=3.7\times 10^{53}\left(\rho^9/E_g^7\right)^{1/2}$ cm$^2$ g$^{-1}$, which is the most 
important absorption opacity of the disk in the super-Eddington regime. 

A pseudo-Newtonian 
potential \citep[][]{PaczynskiWiita1980} is used to mimic the general relativity effects around a Schwarzschild black hole as
\begin{eqnarray}
\phi=-\frac{G\mbh}{r-2r_g},
\end{eqnarray}
where $G$ is the gravitational constant, $\mbh$ is the black hole mass, $r$ is the distance to the central black hole and 
$r_g\equiv G\mbh/c^2$ is the gravitational radius. Notice that the innermost stable circular orbit of this potential is 
$r_{\text{ISCO}}=6r_g$.

We solve the above equations with the new radiation MHD code {\sf Athena++} \citep[][Stone et al. 2017, in preparation]{Stoneetal2008} 
in the spherical polar coordinate system ($r,\theta,\phi$). Logarithmic grid is used for the radial direction so that 
we can cover a large dynamic range. Static mesh refinement is used to increase the resolution 
near the disk midplane but not the polar regions, which significantly reduces the computational cost compared with 
the case without any refinement for the same effective resolution. We refine $r,\theta,\phi$ directions simultaneously 
to achieve the same resolution for all three directions at each level. The fiducial parameters for all the simulations are 
summarized in Table \ref{Table:parameters} and details of the simulation setup are described below. 

\begin{table*}[h]
\caption{Fiducial Simulation Parameters}
\begin{center}
\begin{tabular}{ccc}
\hline\hline
%Variables/Units & {\sf StarB1} & {\sf StarB2} & {\sf StarB3} & {\sf StarB4} & {\sf StarB4z}  \\
Parameters  & Values  & Definition \\
% This corresponds to run StarTopBoxMHD4, StarTopBoxMHD3, StarTopBoxMHD7, StarTopBoxMHD10, StarTopBoxMHD10B
\hline
$\mbh$          &	  $5.00\times 10^8\msun$	    & 	 Black Hole Mass	\\
$r_g\equiv G\mbh/c^2$             &  $7.42\times 10^{13}$ cm & Gravitational Radius\\
$\kappa_{\rm es}$  &    $0.34$ g cm$^{-2}$   &   Electron Scattering Opacity \\
$\Ledd\equiv 4\pi G\mbh c/\kappa_{\rm es}$ &  $7.39\times 10^{46}$ erg s$^{-1}$  &  Eddington Luminosity\\
$\Medd\equiv 10\Ledd/c^2$ &  $8.22\times 10^{26}$ g s$^{-1}$  &  Eddington Accretion Rate\\
$\rho_0$	 &	$1.00\times 10^{-8}$  g cm$^{-3}$	& Fiducial Density \\
$T_0$	 &	$2.00\times10^5$	 K & Fiducial Temperature \\
$P_0$	 &	$2.77\times10^5$ dyn cm$^{-2}$ & Fiducial Pressure \\
$v_0$	 &	$5.26\times10^6$	 cm s$^{-1}$ & Fiducial Velocity  \\
$t_0\equiv r_g/c$	 &	$2.48\times10^3$ s & Fiducial Time \\
$B_0$	 &	$1.87\times10^3$ G & Fiducial Magnetic Field \\
% With the fiducial unit, magnetic pressure is 0.5*P_0, which is x^2/8pi, x in unit of Gauss
%Therefore, B0=(4piP_0)^0.5 in Gauss
\hline
\end{tabular}
\end{center}
\label{Table:parameters}
%\begin{tablenotes}
%\item Note: 
%\end{tablenotes}
\end{table*}

\begin{table*}[h]
\caption{Simulation Parameters}
\begin{center}
\begin{tabular}{ccccc}
\hline
Variables/Units & {\sf AGN150} & {\sf AGN33} & {\sf AGNB25} & {\sf AGNB52}\\
% This corresponds to run AGN4BR, AGNB2, AGNGlobal, AGN4
\hline
$r_i/r_g$              		      		&	  80	    & 	 80		& 	80		&	50	\\	
$\rho_i/\rho_0$			      		&	  50	    &   10		&	10		&	10	\\
$T_i/T_0$                               		&        12.4   &     8.4           &     8.3                   &    8.4            \\
$ \langle P_{r}/ P_{g}\rangle$		&	$7.53\times 10^2$  	    &  	  $3.02\times 10^3$       &	$2.86\times 10^3$		&	$1.81\times 10^4$	\\
$ \langle P_{r}/P_{g}\rangle_{\rho}$   &       $4.94\times 10^2$    	   &     $7.81\times 10^2$             &    $7.40\times 10^2$       &      $7.98\times 10^2$      \\
$ \langle P_{B}/P_{g}\rangle$		 &	 21.55 	    &  	 47.27        &		3.25	&	0.11	\\
$ \langle P_{B}/P_{g}\rangle_{\rho}$   &      $6.77\times 10^{-4}$    		&       $8.83\times 10^{-3}$            &     $8.52\times 10^{-3}$      &    $4.65\times 10^{-2}$        \\
$\Delta r/r$                               		 & 0.024  &	 0.012	&	0.012		&	0.012		\\
$\Delta \theta$		               		&   0.024 &   0.012	&	0.012		&	0.012	 \\
$\Delta \phi$	                       		&  0.024  &  0.012	  &    0.012 	        &	0.012		\\
$N_n$                                     		&  80    &  80  &	80	& 80  \\
$B$ Loops					& Multiple & Multiple   & Single & Single \\
\hline
\end{tabular}
\end{center}
\label{Table:runs}
\begin{tablenotes}
\item Note: The center of the initial torus is located at $r_i$ with density and temperature to be $\rho_i$ and $T_i$. 
For any quantity $a$, $\langle a \rangle$ is the volume averaged value over the torus while $\langle a \rangle_{\rho}$ 
is the averaged value weighted by the mass in each cell. The grid sizes   
$\Delta r, \Delta \theta, \Delta\phi$ are for the finest level at the center of the torus. The number 
of angles for the radiation grid is $N_n$ in each cell. 
\end{tablenotes}
\end{table*}

\section{Simulation Setup}
\label{sec:setup}
We carry out four simulations {\sf AGN150, AGN33, AGNB25} and {AGNB52} to explore the 
properties of super-Eddington accretion disks around a $5\times 10^8\msun$ black hole 
with different accretion rates and magnetic field topologies. 
All the simulations cover a radial range from the inner boundary $4r_g$ to the outer boundary 
$1600r_g$ with $64$ grids, $0$ to $\pi$ along the $\theta$ direction with $32$ grids and 
$0$ to $2\pi$ along $\phi$ direction with $64$ grids in the coarse level. With logarithmic grid 
along the radial direction, we can keep $\Delta r/r=\Delta \theta=\Delta\phi$. We refine the region 
within $20^{\circ}$ along $\theta$ direction on both sides of the disk midplane and radii smaller than $80r_g$ 
with 2 or 3 levels of refinement (see Table \ref{Table:runs}). Periodic boundary conditions are used 
for $\phi$ boundary. For the polar boundary condition, the variables in the ghost zones 
are copied from the last active zones at the same $r$ but with $\phi$ differing by 
$180^{\circ}$. The same mapping is also done for specific intensities propagating along 
the same direction. Vacuum boundary conditions are used for the specific intensities in 
the inner and outer radial boundaries, which means in the ghost zones, 
outgoing specific intensities are copied from the last active zones, while incoming 
specific intensities are set to be zero. All the gas and magnetic field 
quantities are copied from the last active 
zones to the ghost zones except $v_r$, which is only copied if $v_r$ in the last active zones 
points outward. In other cases, $v_r$ is set to be zero in the ghost zones. 

%In this way, we can 
%avoid injecting mass and energy from the radial boundary. 

Initially, we set up a hydrostatic equilibrium 
rotating torus \citep{Hawley2001,Katoetal2004} centered at $r_i=80r_g$, except the run {\sf AGNB25}, where the 
center of the torus is at $r_i=50r_g$ as in \cite{Jiangetal2014c}. 
The radial profile of the 
specific angular momentum of the torus $l$ is assumed to be $l=l_i\left(r\sin\theta/r_i\right)^{0.4}$, 
where $l_i$ is the Keplerian value of the specific angular momentum at $r_i$. With this torus 
shape, the inner edges of the torus are at $40r_g$ and $30r_g$ with $r_i=80r_g$ and $50r_g$ respectively. 
In this way, we do not put any mass inside the region where a steady state disk is going to be formed. 
The density maximum at the center of the torus $\rho_i$ is a free parameter, which determines the 
accretion rate we will get. The larger $\rho_i$ is, the larger the accretion rate we can achieve. 
We also assign a total pressure $\rho_i v_i^2\left(\rho/\rho_i\right)^{\gamma}/\gamma$ with $v_i$ as a 
free parameters, which is then replaced by the sum of gas and radiation pressure by assuming thermal equilibrium. 
The density and temperature at the center of the torus, along with
the averaged ratios between radiation and 
gas pressure for the four runs are summarized in Table \ref{Table:runs}. 
Specific intensities are initialized isotropically in the torus as in \cite{Jiangetal2014c}, 
which adjust within a few steps to give the correct flux. Random 
perturbations with $1\%$ amplitude are added to the density to seed the MRI turbulence. 
Outside the torus, all the variables are set to the floor values with density floor $10^{-8}\rho_0$, temperature 
floor $10^{-2}T_0$ and specific intensity floor $10^{-8}a_rT_0^4$.

Two different initial magnetic field topologies are used in the four simulations. For runs 
{\sf AGNB25} and {\sf AGNB52}, we assume  $\phi$ component of the vector potential 
is proportional to the density while other components are zero. Magnetic fields are then initialized 
with the vector potential, which have the shape of a single loop in the torus. When the inner part of 
the torus is accreted towards the black hole in this case, all the magnetic fields will have the same sign 
of $B_{\theta}$ and net magnetic flux through the disk is expected. For runs {\sf AGN150} and {\sf AGN33}, 
the vector potential is constructed in the similar way as in the Appendix A of \cite{Pennaetal2013b} to create multiple 
loops of magnetic fields in the torus. We construct two loops with the inner edge of the magnetized 
region at $r_i-20r_g$ while the outer edge at $r_i+40r_g$. There are two differences compared with 
\cite{Pennaetal2013b}. First, we scale the vector potential to the density instead of internal energy. 
Second, we multiply the vector potential by the factor $\sin(\pi/2-\theta)$ so that the magnetic field loops 
change sign across the disk midplane. Properties of the initial torus and the resolutions for the four runs are summarized 
in Table \ref{Table:runs}.

These simulations are costly, and were run on $512-4096$ nodes of the Mira
supercomputer at ALCF for a total of 35M core hours.  Access to this system was provided by the DOE
INCITE program, which has been essential for this work.
 
\section{Results}
\label{sec:result}
MRI develops in the initial torus and drives the gas flowing towards the central black hole. 
An accretion disk is built up self-consistently inside the inner edge of the torus. 
We do not control the mass accretion rate in this setup. Instead, it is determined by the available 
mass in the torus and mechanisms for angular momentum transfer as calculated by the 
code. The region where a steady state accretion disk is formed will be the focus of our analysis. 

\subsection{Simulation History}
\begin{figure}[htp]
\centering
\includegraphics[width=1.0\hsize]{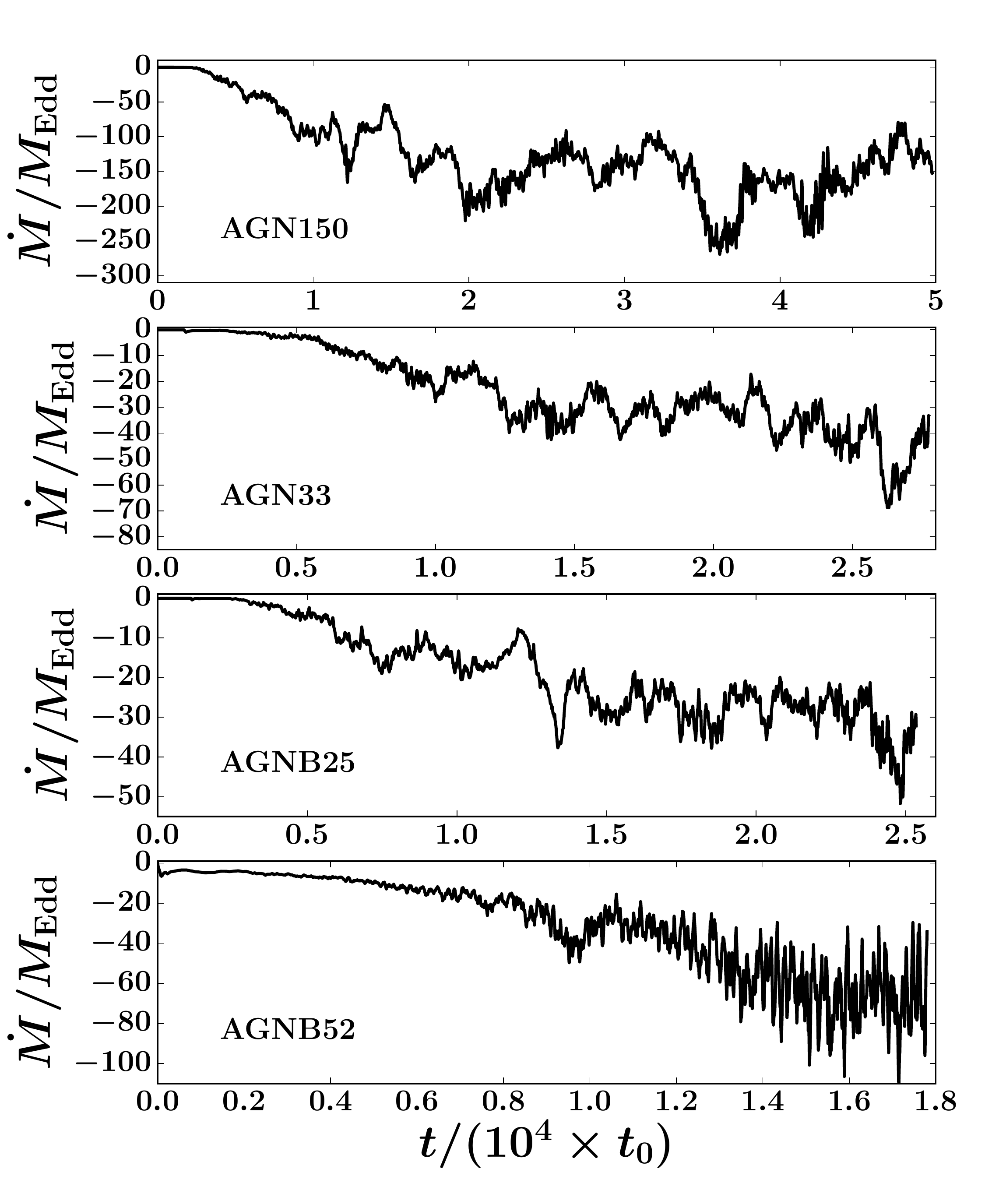}
\caption{Histories of spherically integrated mass accretion rate at $10r_g$ for the four simulations. 
Negative values of $\dot{M}$ mean gas flows towards the black hole. The time 
unit $t_0\equiv r_g/c$ corresponds to $5\times10^{-3}$ Keplerian orbital period at $10r_g$.}
\label{mdot_hist}
\end{figure}

The net mass accretion rate through each radius $r$ can be calculated as
\begin{eqnarray}
\dot{M}=\int_0^{2\pi}\int_0^{\pi} \rho v_r r^2\sin\theta d\theta d\phi.
\label{eqn:mdot}
\end{eqnarray}
Histories of $\dot{M}$ at $10r_g$ for the four simulations 
are shown in Figure \ref{mdot_hist}. After the initial $\sim 10^4t_0$, 
the disks have extended to $10r_g$ with time averaged $\dot{M}$ reaching 
$-146.2,-33.2, -25.8,-51.9\Medd$ for the four runs {\sf AGN150, AGN33, AGNB25, AGNB52} 
respectively. The mass accretion rates show significant fluctuations due to turbulence. 
%, which also show up 
%in the luminosity as discussed in Section XX. 

\begin{figure*}[htp]
\centering
\includegraphics[width=0.48\hsize]{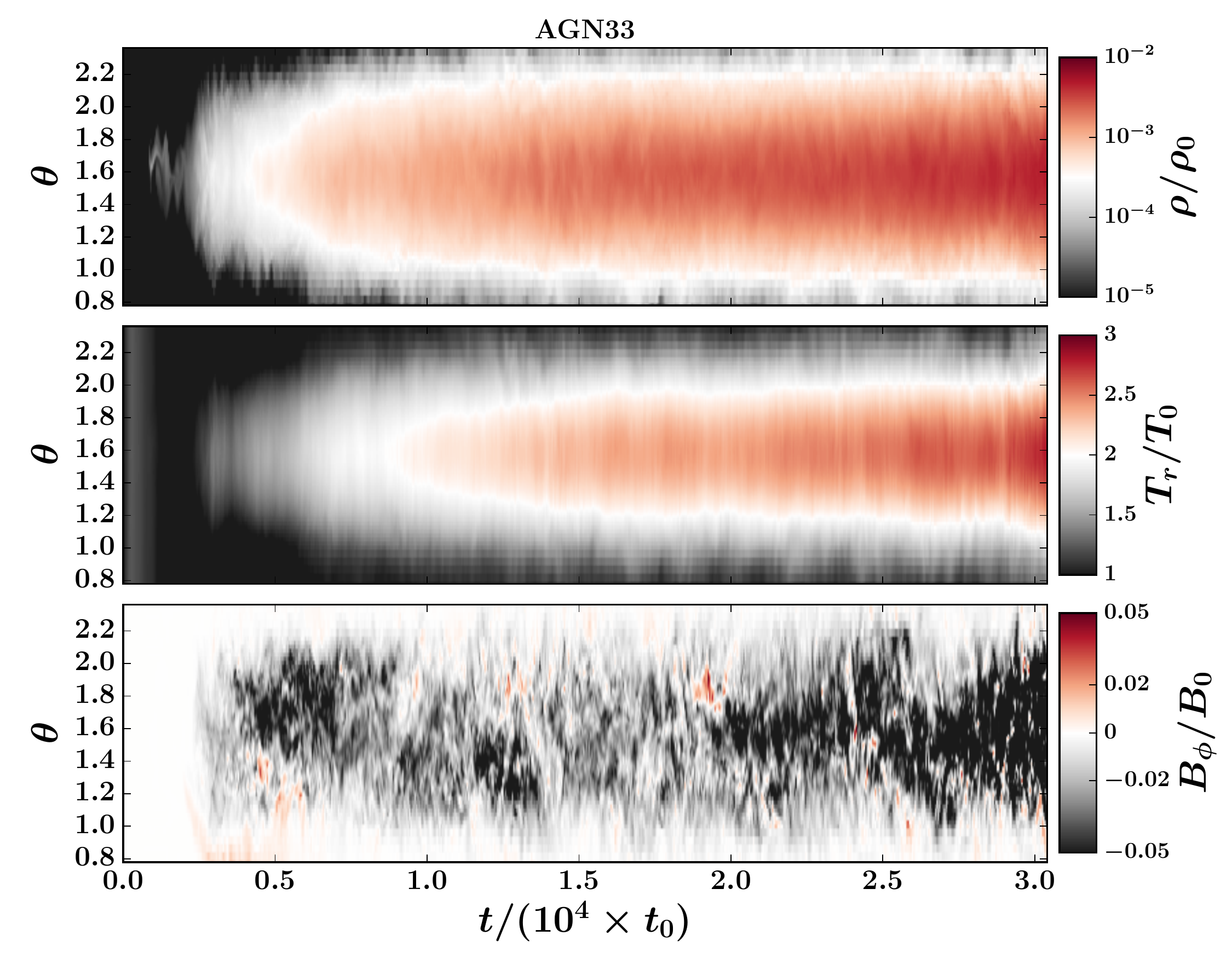}
\includegraphics[width=0.48\hsize]{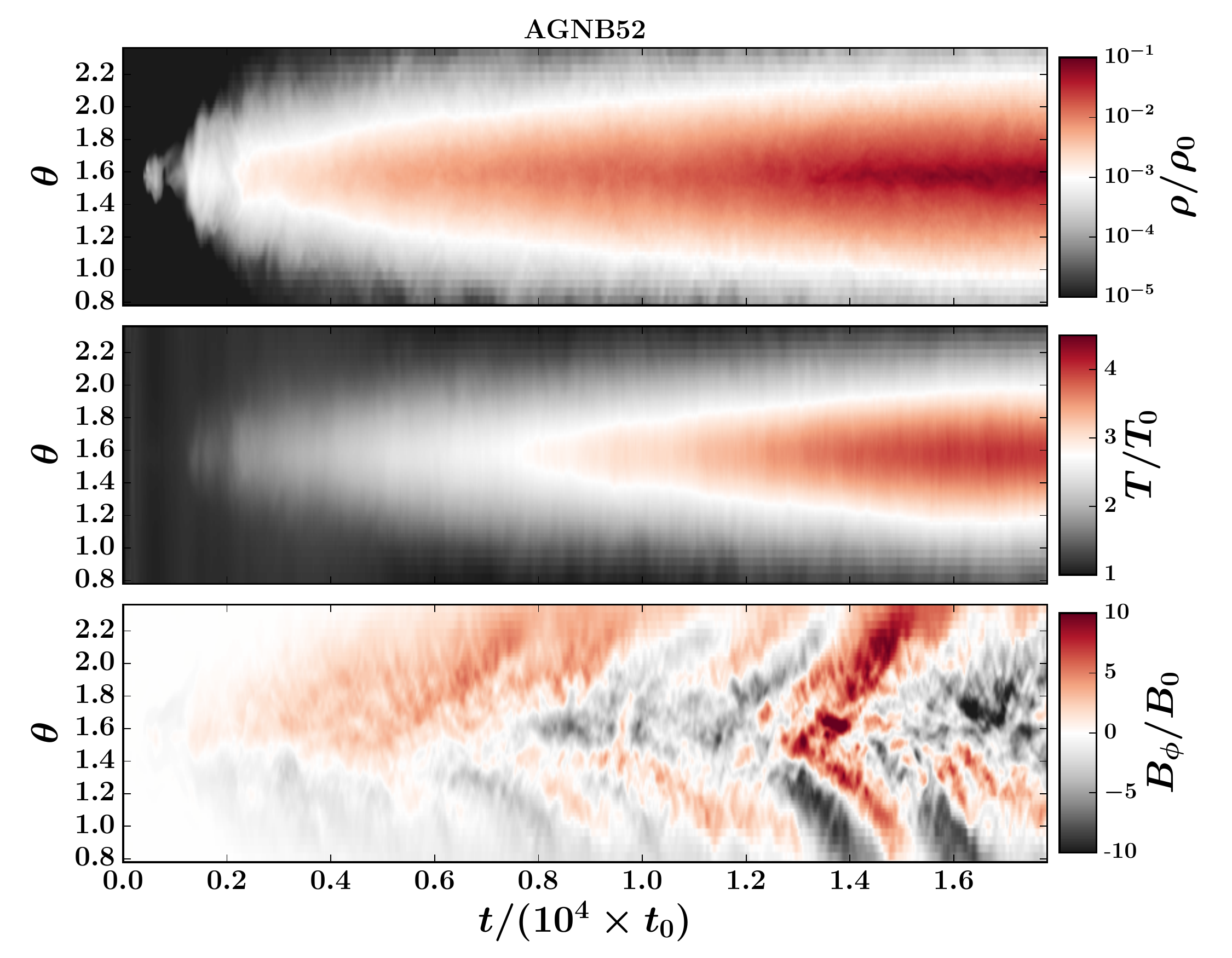}
\caption{Space-time diagram of azimuthally averaged density (top panels), radiation temperature (middle panels) 
  and azimuthal component of magnetic field (bottom panels) at radius $20r_g$ for the run {\sf AGN33} (left) and {\sf AGNB52} (right). (Note that the color bar ranges differ between the left and right
  panels.)
The butterfly diagram shows up in the run {\sf AGNB52} where a single loop of magnetic field is used initially but does not exist 
in {\sf AGN33} which adopts multiple loops of magnetic field in the initial torus.  
}
\label{STplot}
\end{figure*}

All the four runs reach accretion rates significant above $\Medd$ despite different initial magnetic field topologies, although the disks have quite different structures. Histories of 
the azimuthally averaged $\theta$ profiles of density $\rho$, radiation temperature $T_r$ and azimuthal component 
of magnetic field $B_{\phi}$ at $20r_g$ are shown in Figure \ref{STplot} for the two runs {\sf AGN33} and {\sf AGNB52}. 
The space-time diagrams for runs {\sf AGN150} and {\sf AGNB25} are similar to the left and right panels of Figure \ref{STplot} 
respectively. The main differences of the two runs are the evolution of $B_{\phi}$. In the run {\sf AGNB52}, after the disk 
is formed at $20r_g$, $B_{\phi}$ reverses starting near the disk midplane roughly every $4000t_0$, which corresponds 
to $\sim 7$ Keplerian rotation periods at this location. The magnetic fields also rise up from the disk midplane due to 
buoyancy. This is the well known butterfly diagram, which has been observed in previous global simulations with \citep[][]{Jiangetal2014c} 
or without \citep[][]{ONeilletal2011} radiative transfer, as well as various local shearing box simulations of MRI turbulence 
\citep[][]{Stoneetal1996,MillerStone2000,Shietal2010,Davisetal2010,Simonetal2012,Jiangetal2013c,Jiangetal2014b}, although 
the periods for field reversal can differ slightly in these simulations. The butterfly diagram is believed 
to be generated by the dynamo process of MRI \citep[][]{Brandenburgetal1995,Blackman2012} 
and it is an indication that magnetic pressure is amplified so that magnetic buoyancy can play a 
significant role. However, a similar butterfly diagram is not observed in 
the run {\sf AGN33}. Although $B_{\phi}$ shows small scale turbulent structures and 
undergoes vertical oscillations in {\sf AGN33}, we do not see any field 
reversal and systematic buoyantly rising magnetic field from the disk midplane. This suggests that MRI turbulence is 
much weaker in this case and other mechanisms must be responsible for the angular momentum transfer so that the 
simulations {\sf AGN150} and {\sf AGN33} can reach similar accretion rates as in {\sf AGNB25} and {\sf AGNB52}. This will be examined in Section \ref{sec:stress}.

The vertical profiles of density are more concentrated towards the disk midplane in {\sf AGNB52} compared with the density 
profiles in {\sf AGN33} as shown in Figure \ref{STplot}. Particularly, it happens when strong  magnetic fields 
buoyantly rise up from the disk midplane. 
This is consistent with the simulation shown in \cite{Jiangetal2014c} and is closely related to the 
butterfly diagram. The strong magnetic buoyancy in {\sf AGNB52} moves the dissipation away from the disk midplane 
and reduces the disk scale height compared with the case without strong magnetic buoyancy as in {\sf AGN33}. 
The radiation temperature varies from $\sim 10^{5}-10^{6}$ K and the midplane temperature is larger in {\sf AGNB52} because of the larger accretion rate.

\subsection{Spatial Structures of the Disks}
\label{sec:spatial}

\begin{figure}[htp]
\centering
\includegraphics[width=1.0\hsize]{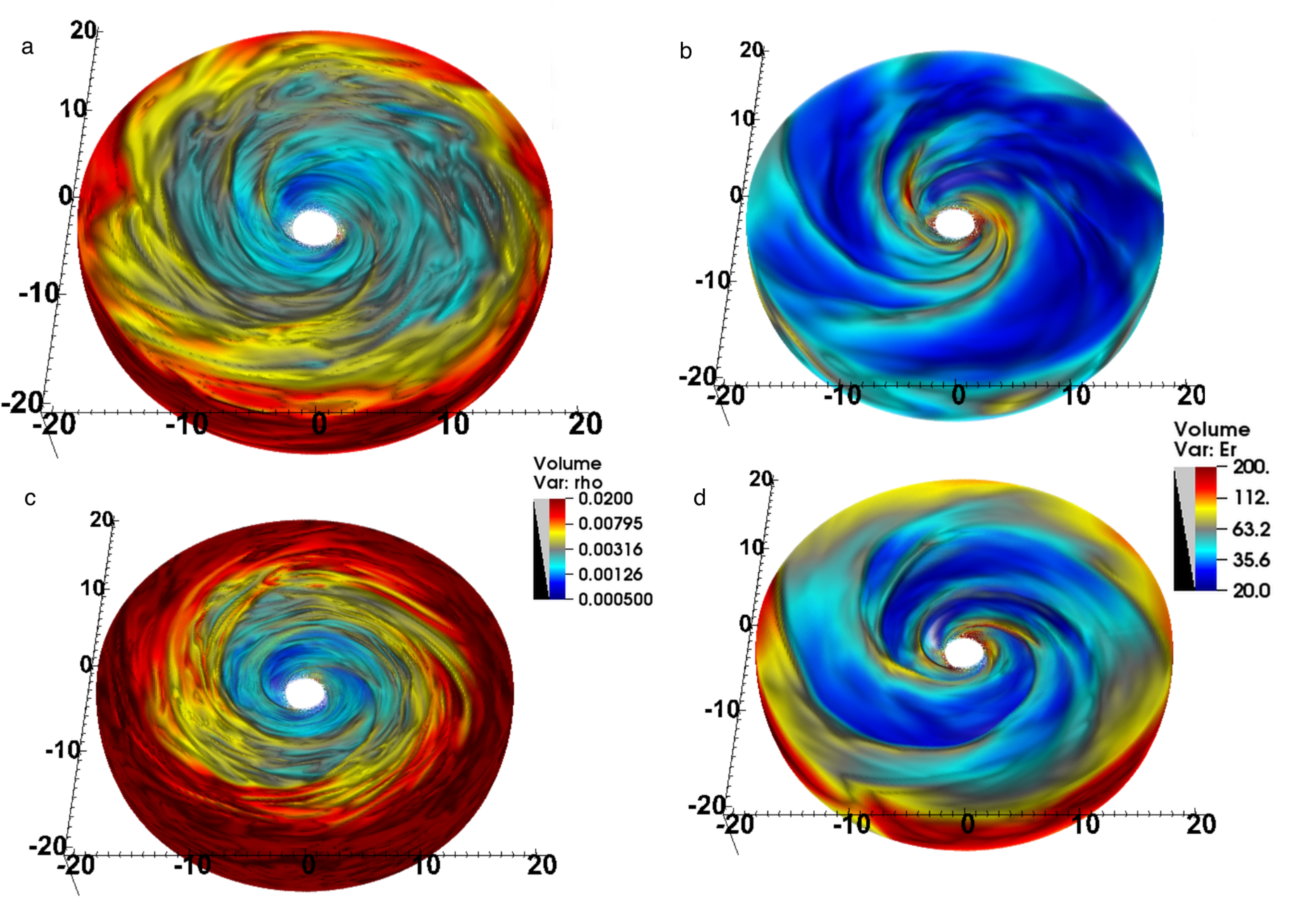}
\caption{Snapshots of density $\rho/\rho_0$ (left panels a and c) and 
radiation energy density $E_r/a_rT_0^4$ (right panels b and d) 
at the inner $40r_g$ of the accretion disks for runs {\sf AGN33} (top panels a and b) 
and {\sf AGNB25} (bottom panels c and d). The length unit in the plot is $2r_g$. 
The snapshots are taken at times $2.73\times 10^4t_0$ 
and $2.53\times 10^4t_0$ for runs {\sf AGN33} and {\sf AGNB25} respectively. 
Notice the significant spiral arms in the disk, particularly for $E_r$.}
\label{spiralimage}
\end{figure}

Snapshots of density $\rho$ and radiation energy density $E_r$ at 
the inner regions of the accretion disks for {\sf AGN33} and {\sf AGNB25} 
are shown in Figure \ref{spiralimage}. In both cases, the density structures are fully turbulent 
as in previous 3D global accretion disk simulations with MRI turbulence \citep[][]{Armitage1998,Hawley2001,Jiangetal2014c}.
The striking features are the spiral arms in the snapshots of radiation energy density, which also 
show up in the snapshots of density in spite of the turbulence. The spiral structures exist 
in all the four runs and they cause significant azimuthal variations of density and temperature.  Excitations 
of density waves and the associated spiral shocks by the MRI turbulence 
have been observed in previous local \citep[][]{GardinerStone2005} and global \citep[][]{Armitage1998} simulations 
and have been suggested to be caused by the generic swing application process driven by the potential vorticity \citep[][]{HeinemannPapaloizou2009a,
HeinemannPapaloizou2009b}. This is the first time that the development of coherent global density waves and spiral shocks have been seen in simulations with self-consistent thermodynamics in the strongly radiation pressure dominated AGN accretion disks. Unlike the weak spiral shocks and small 
associated Reynolds stress found in previous calculations, we will show in Section \ref{sec:stress} that the density waves generate 
very strong shocks in our simulations and result in very large Reynolds stress.

\begin{figure}[htp]
\centering
\includegraphics[width=1.0\hsize]{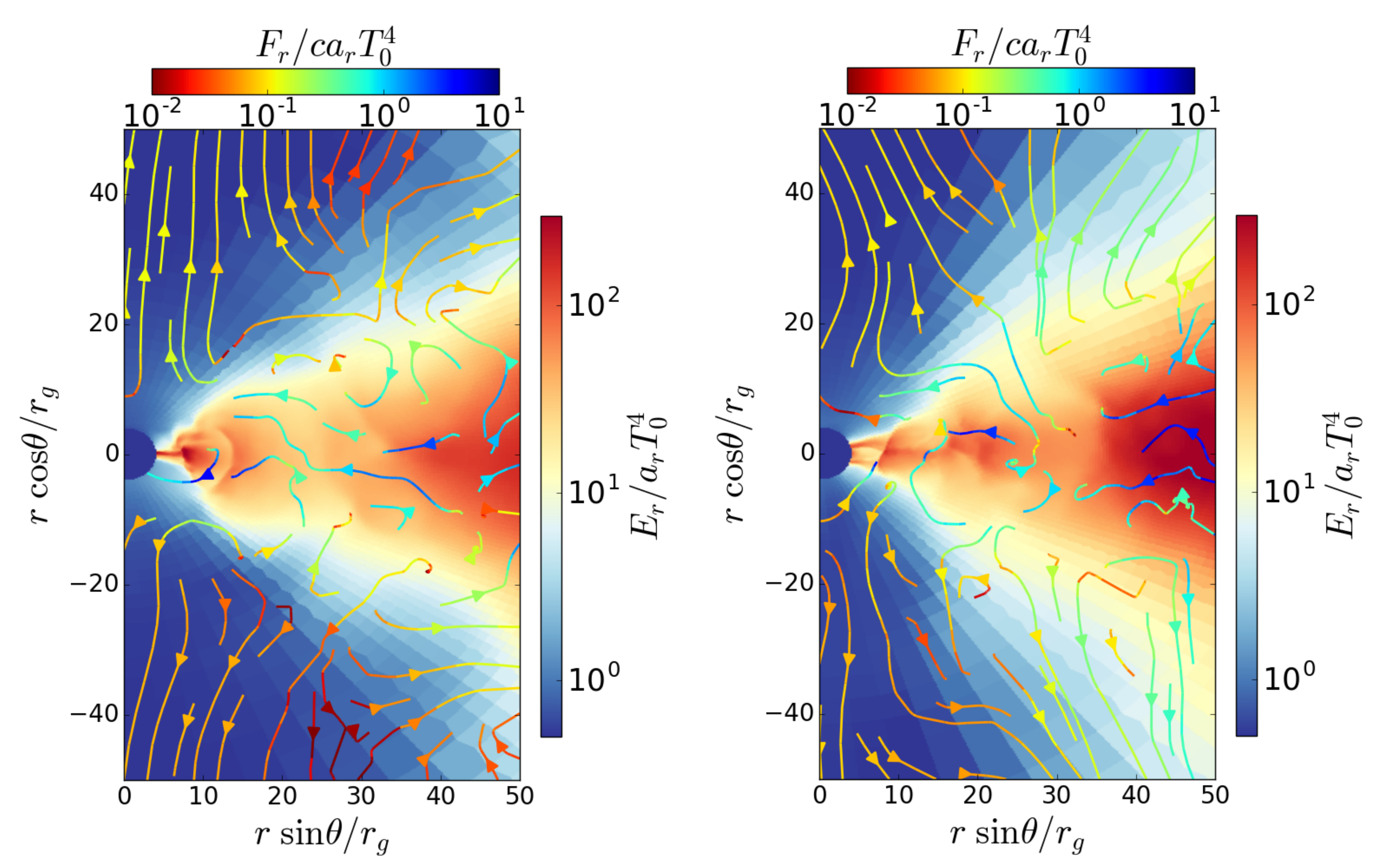}
\caption{Slices of radiation energy density $E_r$ (color) and 
$r,\theta$ components of radiation flux $F_{r,r},\ F_{r,\theta}$ (streamlines) through the plane $\phi=\pi$ 
at the same times as in Figure \ref{spiralimage} for the two runs 
{\sf AGN33} (left panel) and {\sf AGNB25} (right panel).  }
\label{spiralslice}
\end{figure}

Slices of $E_r$ as well as radial and poloidal components of radiation flux $F_{r,r}, 
F_{r,\theta}$ through the plane $\phi=\pi$ at the same times are shown in Figure \ref{spiralslice}. 
Locations of sharp $E_r$ jumps are caused by the spiral shocks, which extend to large angles 
along the $\theta$ direction. Multiple shock positions correspond to the spiral arms shown in Figure \ref{spiralimage}. 
The lab frame radiation flux is dominated by the advection component $\bv E_r$ in the main body of the disk. 
Then photons leave the disk through the low density funnels.

\begin{figure*}[htp]
\centering
\includegraphics[width=1.0\hsize]{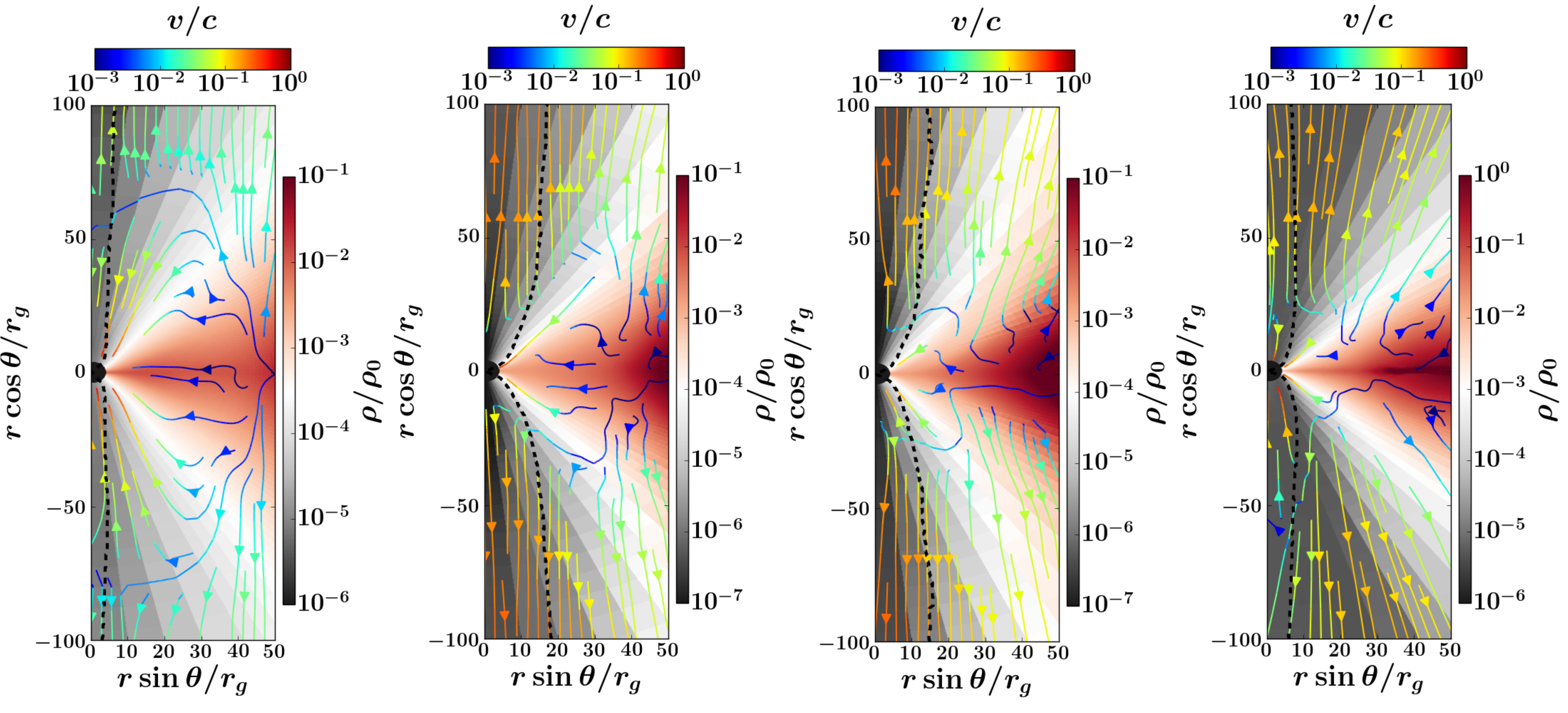}
\caption{Time and azimuthally averaged spatial structures of density 
$\rho$ (color) and density weighted flow velocity $v_r, v_{\theta}$ (streamlines)
at the inner regions of the disks. Color of the streamlines represents the 
velocity magnitude $v\equiv \sqrt{v_r^2+v_{\theta}^2}$. From left to right, they are for runs 
{\sf AGN150, AGN33, AGNB25, AGNB52} respectively. The dashed black lines 
indicate the locations where the integrated optical depth from the rotation axis along the
horizontal direction is one. }
\label{averhov}
\end{figure*}

\begin{figure*}[htp]
\centering
\includegraphics[width=1.0\hsize]{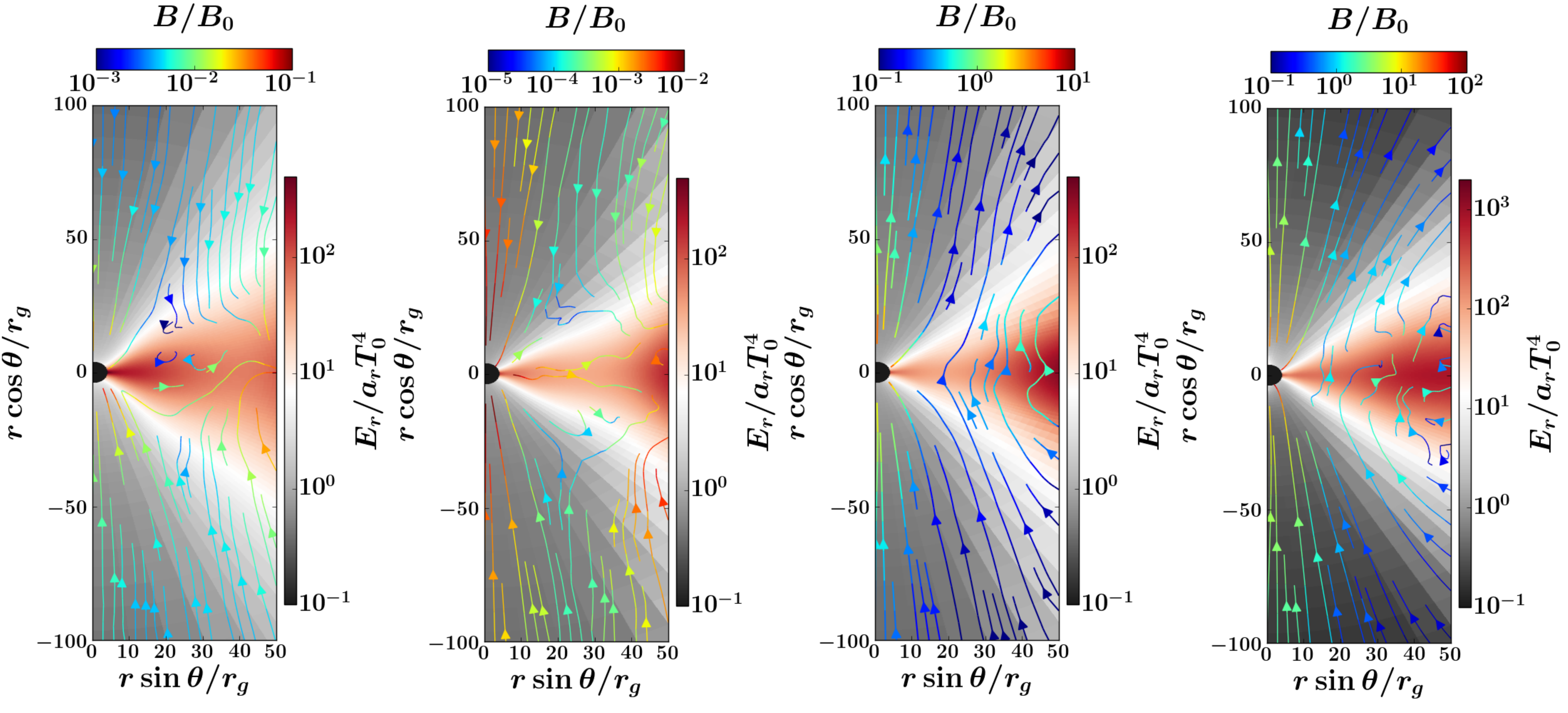}
\caption{Time and azimuthally averaged spatial structures of radiation 
energy density $E_r$ (color) and magnetic fields $B_r,\ B_{\theta}$ (streamlines). 
Color of the streamlines represent $B\equiv \sqrt{B_r^2+B_{\theta}^2}$. 
From left to right, they are for runs {\sf AGN150, AGN33, AGNB25, AGNB52} 
as in Figure \ref{averhov}. The time average is also done during the same time 
as in that Figure. There are net vertical magnetic fields through the disk 
in the third and fourth panels but not in the first two panels. }
\label{aveErB}
\end{figure*}

Time averaged spatial structures of the disks are studied during the periods 
$2.05\times 10^4-4.98\times 10^4 t_0$,  $1.71\times 10^4-2.78\times 10^4 t_0$,  $1.71\times 10^4-2.54\times 10^4 t_0$,  $1.37\times 10^4-1.78\times 10^4 t_0$ for runs {\sf AGN150,AGN33,AGNB25} and {\sf AGNB52} respectively, which are 
chosen when the inner regions of the disks have reached an approximately constant $\dot{M}$ (see Section \ref{sec:radial}). 
This is also the time range we will average the data for the following analysis.  
We calculate the time and azimuthally 
averaged  $\rho,\rho v_r,\rho v_{\theta}, E_r, B_r, B_{\theta}$, where $v_r, B_r$ and $v_{\theta},B_{\theta}$ 
are the radial and poloidal components of the flow velocity and magnetic fields.
The averaged momentum is then divided by the averaged density to get the density weighted 
flow velocity. These averaged quantities at the inner region of the disks for the four runs are shown in Figure \ref{averhov} 
and \ref{aveErB}. The spiral shock structures do not show up after the azimuthal average. Gas flows 
towards the black hole near the disk midplane while strong outflow with velocity $\sim 0.1-0.4c$ develops 
along the low density funnels. Despite the different magnetic field configurations, the runs {\sf AGN33} and 
{\sf AGNB25} show very similar time averaged structures when they reach similar accretion rates. Magnetic 
pressure in the funnel region is much stronger in runs {\sf AGNB25} and {\sf AGNB52} when there are net 
poloidal magnetic field lines through the disks. %, which reduces the density in this region for a given accretion rate. 
The stagnation points in the funnels, where the velocities change from inflow to outflow, are located at 
$\sim 10-20r_g$, except for the run {\sf AGN150}. 
When $\dot{M}$ reaches $\sim 200\Medd$, 
the stagnation points move to $\sim 50-60r_g$ and all the funnel becomes optically thick in this region. 
As outflow is now launched from a much larger  radius, the outflow velocity is actually smaller in this case. Properties 
of the outflow will be quantified in Section \ref{sec:outflow}. 

\subsection{Radial Profiles of the Disks}
\label{sec:radial}
For any quantity $a$, we calculate the time and shell averaged value at each radius $r$ as
\begin{eqnarray}
\<a\>=\frac{1}{4\pi\Delta t}\int_0^{2\pi}\int_0^{\pi}\int_{t_1}^{t_2} a\ dt\sin\theta d\theta d\phi,
\end{eqnarray}
where $\Delta t=t_2-t_1$ with the time interval $t_1, t_2$ as specified in Section \ref{sec:spatial}. 
The mass weighted average is calculated as
\begin{eqnarray}
\<a\>_{\rho}=\frac{1}{4\pi \Delta t\<\rho\>}\int_0^{2\pi}\int_0^{\pi}\int_{t_1}^{t_2} a\rho\ dt\sin\theta d\theta d\phi.
\end{eqnarray}

\begin{figure}[htp]
	\centering
	\includegraphics[width=1.0\hsize]{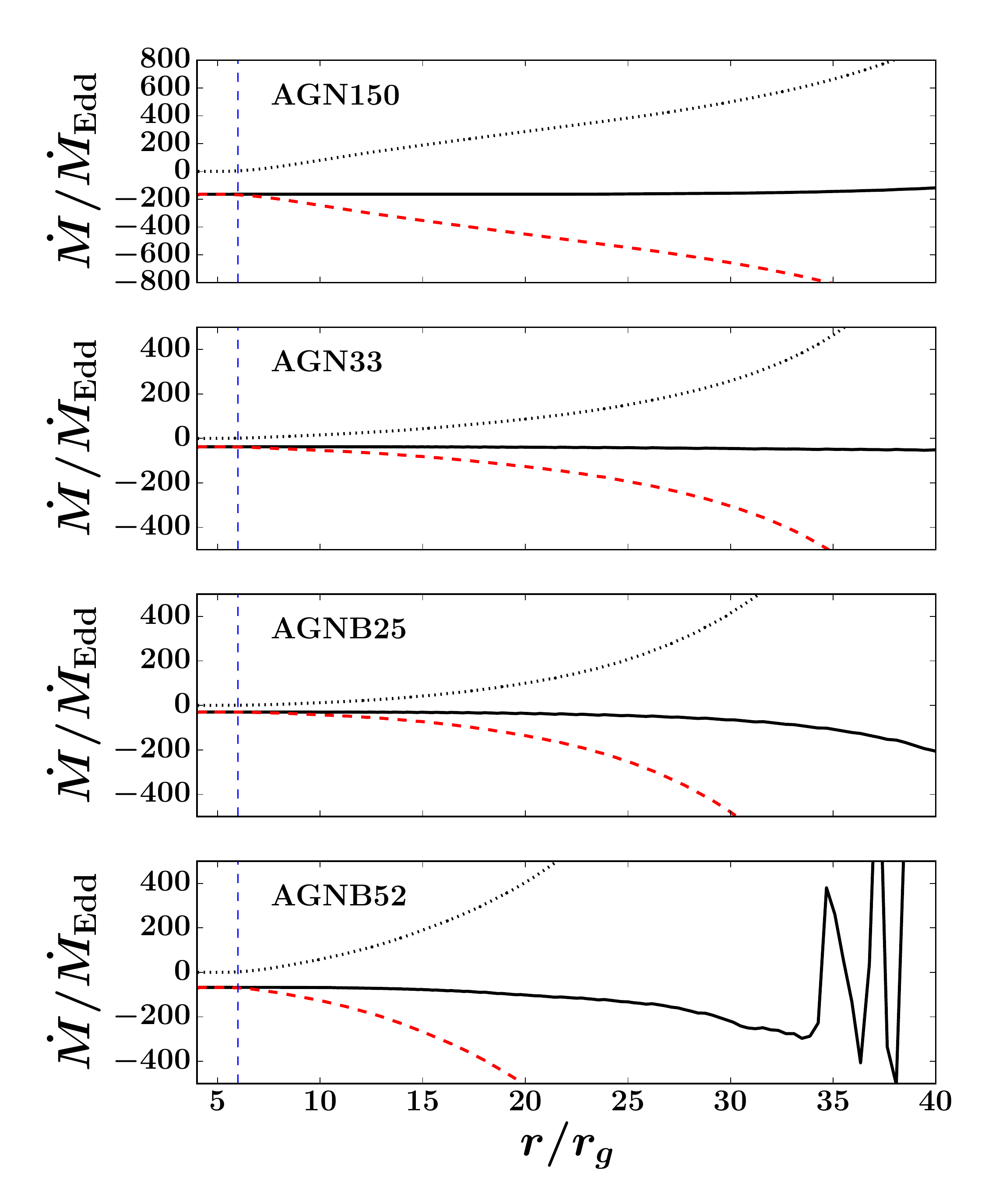}
	\caption{Time and shell averaged radial profiles of mass accretion rate $\dot{M}$ in unit 
		of the Eddington mass accretion rate $\dot{M}_{\rm Edd}$ defined in Table \ref{Table:parameters}. 
		From top to bottom, they are for the four runs  {\sf AGN150, AGN33, AGNB25, AGNB52}. In each panel, 
		the solid black lines are the net mass accretion rates $\dot{M}$, which are $-150, -33, -25, -52\dot{M}_{\rm Edd}$ 
		in the steady state inner regions of the disks. The dotted black and dashed red lines are the mass flux carried by 
		the gas with $v_r>0$ and $v_r<0$ respectively. The dashed vertical lines indicate the location of ISCO ($6r_g$).}
	\label{avemdot}
\end{figure}

Radial profiles of the average net mass accretion rates $\<\dot{M}\>$ (equation \ref{eqn:mdot}) for the four runs 
{\sf AGN150, AGN33, AGNB25, AGNB52} are shown in Figure \ref{avemdot}. The net mass accretion rates roughly 
reach constant values up to $35r_g$ for {\sf AGN150, AGN33}, $25r_g$ for {\sf AGNB25} and $20r_g$ for {\sf AGNB52} 
during the time we have run these simulations, which indicate steady state disks in these regions. 
The net mass accretion rates can be decomposed into 
two components with $v_r>0$ and $v_r<0$, which are also shown in Figure \ref{avemdot} for comparison
with the net accretion rate. The mass fluxes associated with the ingoing
and outgoing components are much larger than $\dot{M}$, but this is predominantly due to turbulent
motions in the flow. In particular, the outgoing
component always begins outside ISCO, as expected, and increase rapidly with radius. We emphasize
that only a small fraction of this outgoing component is associated with an unbound outflow, which will be quantified in 
Section \ref{sec:outflow}. It is interesting to notice that these inflow and outgoing components are much larger than the stellar mass black hole case for a similar net mass accretion rate $\dot{M}/\dot{M}_{\rm Edd}$ (Figure 4 of \citealt{Jiangetal2014c}).

\begin{figure*}[htp]
	\centering
	\includegraphics[width=0.45\hsize]{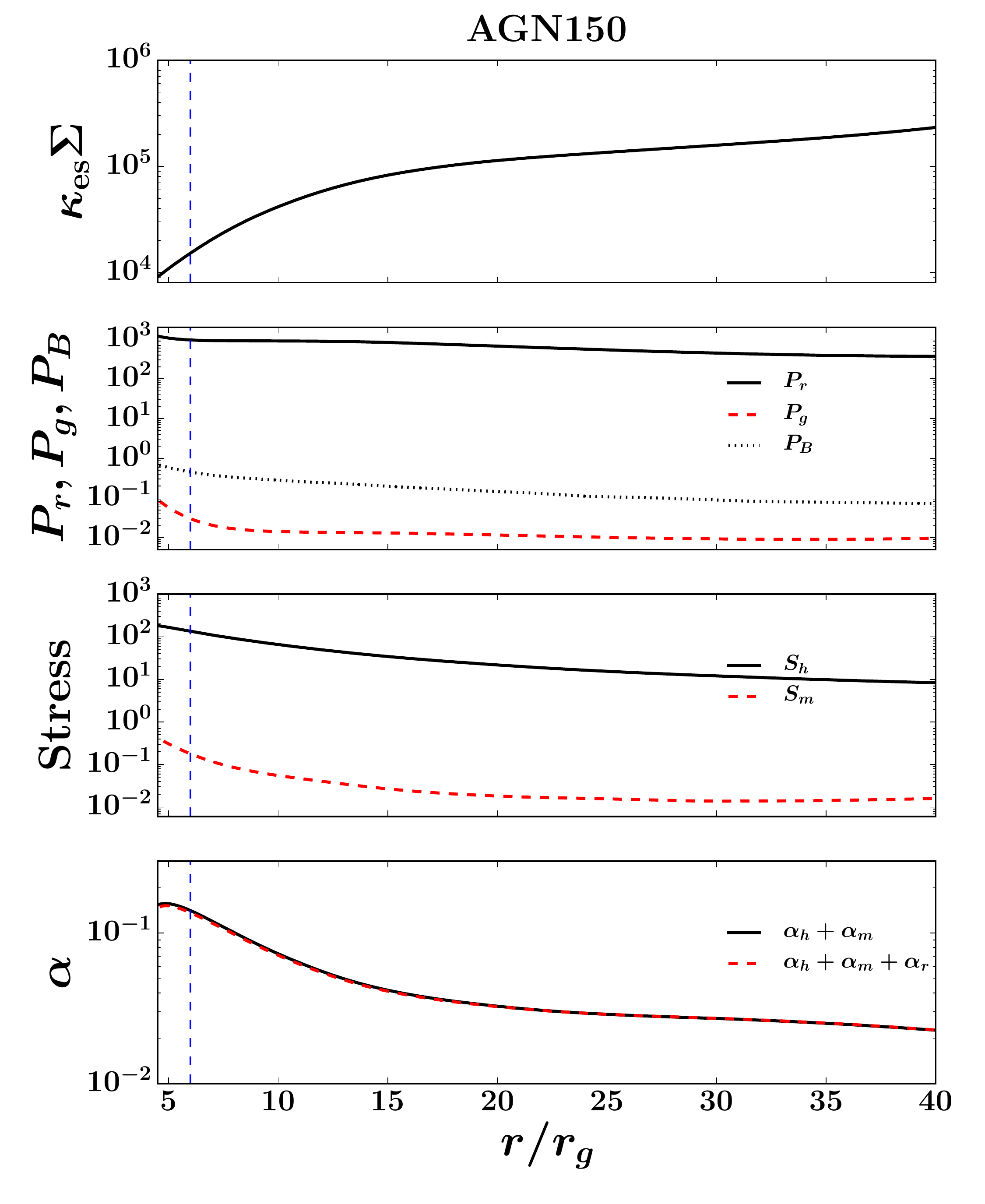}
	\includegraphics[width=0.45\hsize]{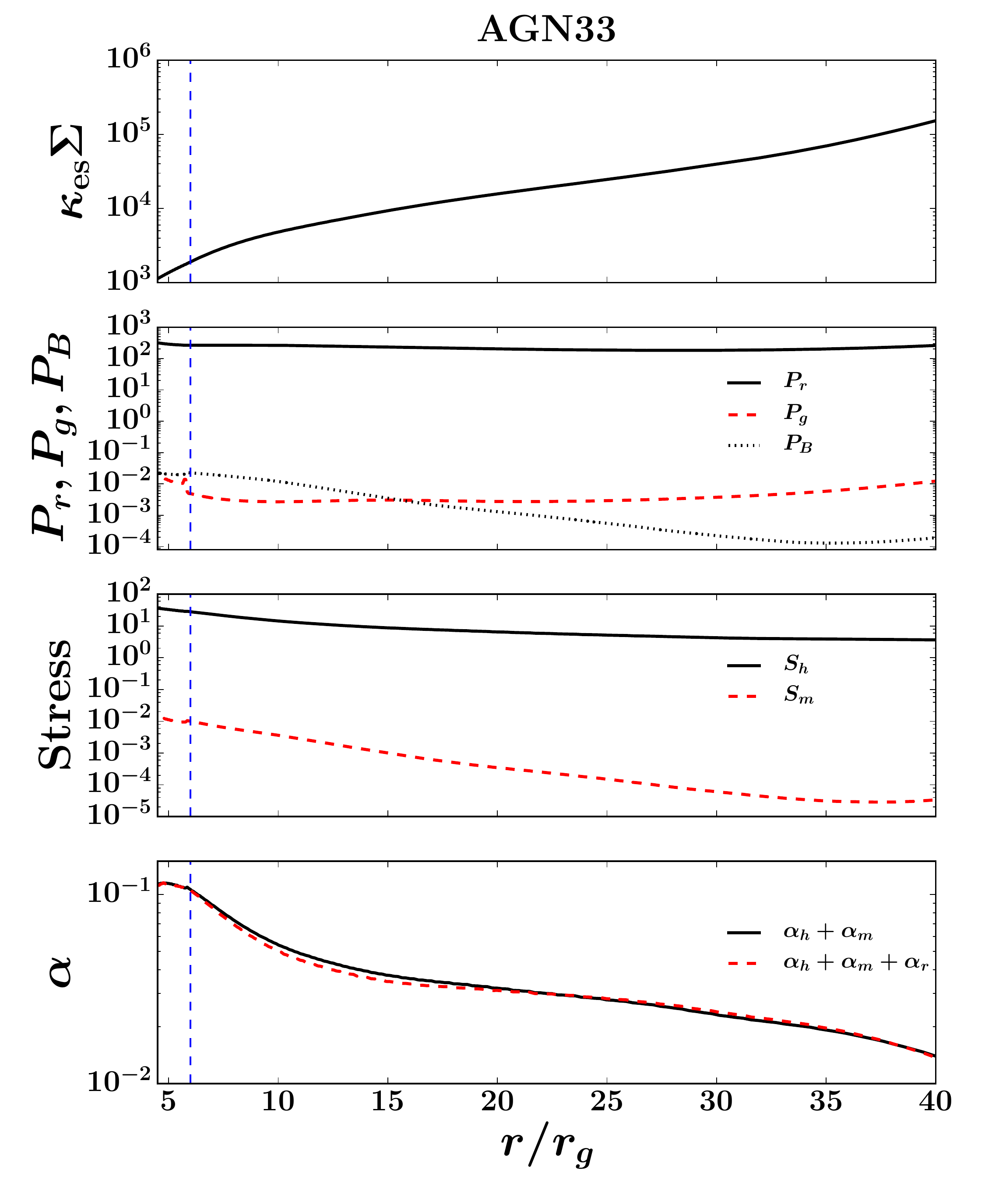}
	\includegraphics[width=0.45\hsize]{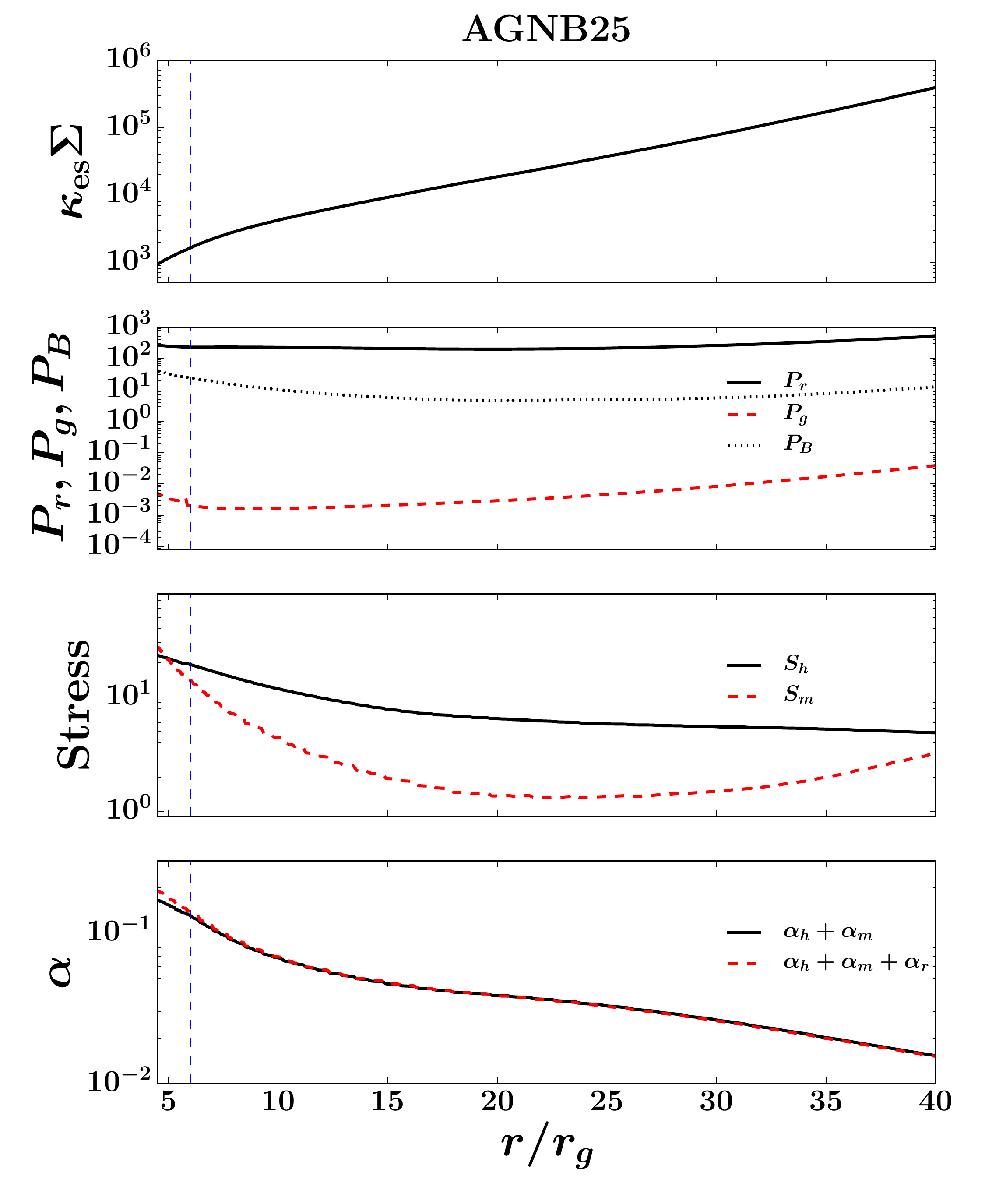}
	\includegraphics[width=0.45\hsize]{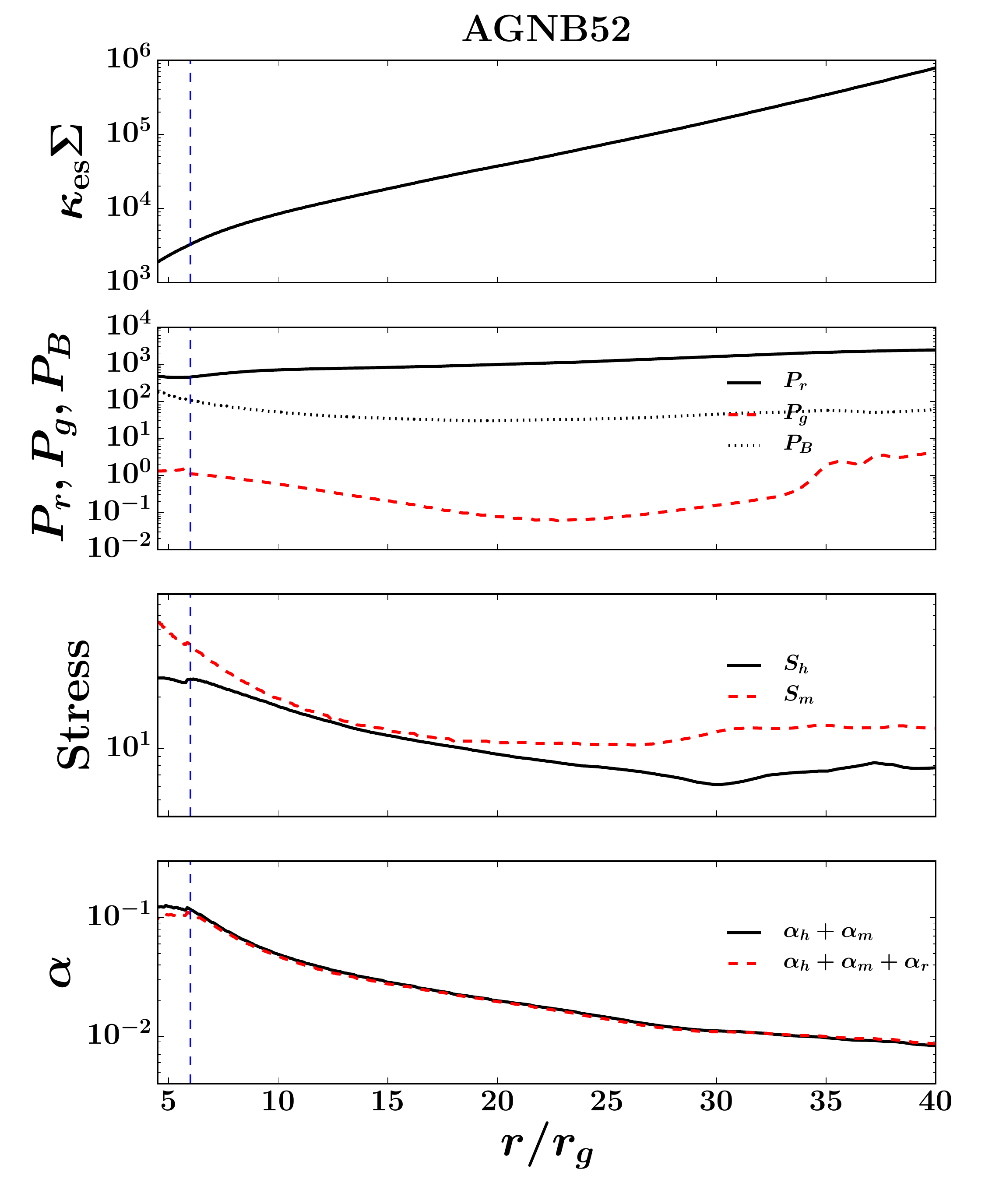}
	\caption{Time and shell averaged radial profiles of various quantities as described in Section \ref{sec:radial} for the four runs 
		as indicated in the figures. From top to bottom panels of each plot, they are surface 
		density multiplied by the electron scattering opacity $\kappa_{\rm es}$, radiation ($P_r$, black solid lines), 
		gas ($P_g$, red dashed lines) and magnetic ($P_m$, black dotted lines) pressure, Reynolds ($S_h$, black solid lines) and Maxwell ($S_m$, red dashed lines) stress, 
		effective $\alpha$ for Reynolds and Maxwell stress alone ($\alpha_h+\alpha_m$, black solid lines) as well as effective $\alpha$ including the radiation viscosity ($\alpha_h+\alpha_m+\alpha_r$,red dashed lines). The vertical dashed lines show the location of ISCO. Pressure and stress have the fiducial units given in Table \ref{Table:parameters}.}
	\label{fig:radial}
\end{figure*}

%sigma \propto r^1.8 for AGNM2, r^1.814 for AGNM1, R^2.028 for AGNBM1, r^2.751 for AGNBM2

The surface density $\Sigma$ at each radius can be calculated as $\Sigma=\int_0^{2\pi}\int_0^{\pi}\rho r\sin\theta d\theta d\phi$. The time averaged 
radial profiles of $\Sigma$ multiplied by the electron scattering opacity $\kappa_{\rm es}$ are shown in the top panels of Figure \ref{fig:radial} 
for the four simulations. The total optical depth is always larger than $10^3$ and the largest in {\sf AGN150}, corresponding to the largest accretion rate. 
In each simulation, surface density increases with radius as $r^{1.8}$ for {\sf AGN150, AGN33},
$r^{2.0}$ for {\sf AGNB25} and $r^{2.8}$ for {\sf AGNB52}, which is consistent with the density profiles shown in \cite{Jiangetal2014c}. 
Similar to the stellar mass black hole case, the effective absorption opacity in the inner regions of these disks is much smaller and  approaches optically thin regime. Gas and radiation temperature can differ slightly even near the disk midplane as shown in Section \ref{sec:vertical}. 

The time and shell averaged radial profiles of isotropic radiation pressure ($P_r=E_r/3$), gas pressure $P_g$ and magnetic pressure $P_B$ are shown in the second panels of Figure \ref{fig:radial}. The ratio between radiation and gas pressure varies from $10^3$ to $10^5$, which is significantly larger than 
the ratio in the stellar mass black hole case as shown in Figure 9 of \cite{Jiangetal2014c}. For runs {\sf AGN150} and {\sf AGN33} where there are no net vertical magnetic flux through the disk, radiation pressure is larger than the magnetic pressure by a factor of $10^4-10^5$. However, with net vertical magnetic fields threading the disk as in runs {\sf AGNB25} and {\sf AGNB52}, the ratio between radiation and magnetic pressure varies from 
$5$ to $50$, which is consistent with the values found in previous local shearing box and global simulations of MRI turbulence in radiation pressure dominated flows when the radiation pressure is only a factor of a few larger than the gas pressure \citep[][]{Hiroseetal2009,Jiangetal2013c,Jiangetal2014c}.

Stress responsible for the angular momentum transfer in the disk comes from Maxwell stress $S_m$, Reynolds stress $S_h$ and radiation stress 
$S_{rad}$. In spherical polar coordinate, we calculate the radial profiles of the volume and time averaged $r-\phi$ stress as $S_m=\<-B_rB_{\phi}\>$, $S_h=\<\rho v_r v_{\phi}\>-\<\rho v_r\>\<v_\phi\>$. Here for the Reynolds stress, we subtract the angular momentum flux carried by the mean inflow in the disk. Similarly, the off-diagonal component of radiation pressure in the lab-frame $P_r^{r\phi}$ includes contributions from radiation viscosity in the co-moving frame as well as the inflow angular momentum flux carried by the radiation when photons are advected inwards with the mean accretion flow. This contributions is included by the  Lorentz transformation of the specific intensities due to the azimuthally averaged radial inflow $\langle v_r\rangle$ and rotation velocities $\langle v_{\phi}\rangle$ (see discussions in Section \ref{sec:radvis}). To subtract this component, we transform specific intensity $E_{r,0}/4\pi$ for each angle from the co-moving frame to the lab-frame with velocity $\langle v_r\rangle$ and $\langle v_{\phi}\rangle$, where $E_{r,0}$ is the radiation energy density in the co-moving frame. The off-diagonal component of the second moment of these new lab-frame intensities is then subtracted from $P_r^{r\phi}$, the result of which is $S_{rad}$. 
The corresponding effective $\alpha$ values can be defined as $\alpha_m\equiv S_m/\left(\<P_r\>+\<P_g\>+\<P_B\>\right)$, 
$\alpha_h\equiv S_h/\left(\<P_r\>+\<P_g\>+\<P_B\>\right)$, $\alpha_r\equiv S_{rad}/\left(\<P_r\>+\<P_g\>+\<P_B\>\right)$. Radial profiles of the averaged stresses and effective $\alpha$ values for the four runs are shown in Figure \ref{fig:radial}. For the two runs with multiple magnetic field loops 
({\sf AGN150} and {\sf AGN33}), Reynolds stress is actually larger than the Maxwell stress by a factor of $10^3$. The Maxwell stress increases significantly when a single loop of magnetic fields is used as in {\sf AGNB25} and it only becomes larger than the Reynolds stress in the run {\sf AGNB52} when the net poloidal magnetic flux through the disk is increased by a factor of $\sim 2-4$ in the inner region of the disk. 
The relative small Maxwell stress in {\sf AGN150} and {\sf AGN33} is also consistent 
with the small magnetic pressure in the two runs, as the ratio between $\<-B_rB_{\phi}\>$ and $\<P_B\>$ is always $\sim 0.2-0.4$ in these simulations 
as found in most simulations with MRI turbulence \citep[][]{Blackmanetal2008,Guanetal2009,Sorathiaetal2012,Jiangetal2013b,Hawleyetal2011,Hawleyetal2013,Jiangetal2014c}. We have also checked that in all the four runs, the total magnetic pressure is dominated by the turbulent magnetic pressure with contributions from the mean magnetic fields component less than $10\%$. The time and volume averaged radial 
component of magnetic fields $B_r$ is almost zero in all the four runs, which confirms that the Maxwell stress we get from these simulations are indeed due to the turbulence instead of the stress caused by the mean magnetic field components $<B_r>$ and $<B_{\phi}>$. 

The increased saturation level of Maxwell stress with net vertical magnetic flux through the disk has been found in various local shearing box as 
well as global simulations of MRI turbulence \citep[][]{HGB1995,BaiStone2013,Fromangetal2013,Simonetal2013,Bethuneetal2016,ZhuStone2017}. However, in most of these ideal MHD simulations with isothermal equation of state, Maxwell stress is always larger than the Reynolds stress by a factor of $4-5$. The ratio between Maxwell stress and total pressure is also larger than $\sim 1-3\%$ in both cases with or without net vertical magnetic flux. 
For simulations of MRI turbulence in the radiation pressure dominated flows when the radiation pressure is only larger than the gas pressure by a factor of $\sim 10$ and radiation sound speed is still much smaller than the speed of light \citep[][]{Hiroseetal2009,Blaesetal2011,Jiangetal2013c,Jiangetal2014c}, similar results are found for the properties of MRI turbulence. However, \cite{Jiangetal2013b} studied the saturation of MRI turbulence in strongly radiation pressure dominated regime based on unstratified local shearing box simulations and found that when radiation pressure was more than $100$ times larger than the gas pressure, the ratio between Maxwell stress and total pressure became smaller than $1\%$. This is likely caused by the damping due to radiation viscosity in the very compressible radiation pressure dominated flow and consistent with the results shown in Figure \ref{fig:radial}. In these local shearing box simulations, Reynolds stress is always smaller than the Maxwell stress and the ratio between Maxwell stress and Reynolds stress becomes larger in the strongly radiation pressure dominated regime. 
This is different from what we find in the four simulations. As we will show in Section \ref{sec:stress}, the large Reynolds stress in these global simulations is produced by the density waves as shown in Figure \ref{spiralimage}, which are much weaker in local shearing box simulations. 
%discuss the resolution convergence study at the end

The sum of $\alpha_h,\alpha_m$ and $\alpha_r$ are almost the same as $\alpha_h+\alpha_m$ as shown at the bottom panels of Figure \ref{fig:radial} for all the four runs, which means radiation stress is much smaller than Maxwell and Reynolds stresses, even though the radiation pressure is much larger than the gas pressure. This is because the anisotropic component of the co-moving radial field is significantly reduced by the small mean free path to electron scattering. %, which will be quantified in Section XX. 
Even the relative contributions from Maxwell and Reynolds stresses vary significantly among the four runs, the ratio between the total stress and pressure is actually very similar. The effective $\alpha$ varies from $\sim 0.02$ at the outer edge of the steady state disk to $\sim 0.1$ at the inner edge of the simulation domain for all the four runs as shown in the bottom panels of Figure \ref{fig:radial}, which is also similar to the profile we get for the stellar mass black hole case \citep[][]{Jiangetal2014c}. This is perhaps why all these disks can reach such high accretion rates despite different mechanisms controlling the transfer of angular momentum.

\begin{figure}[htp]
	\centering
	\includegraphics[width=0.9\hsize]{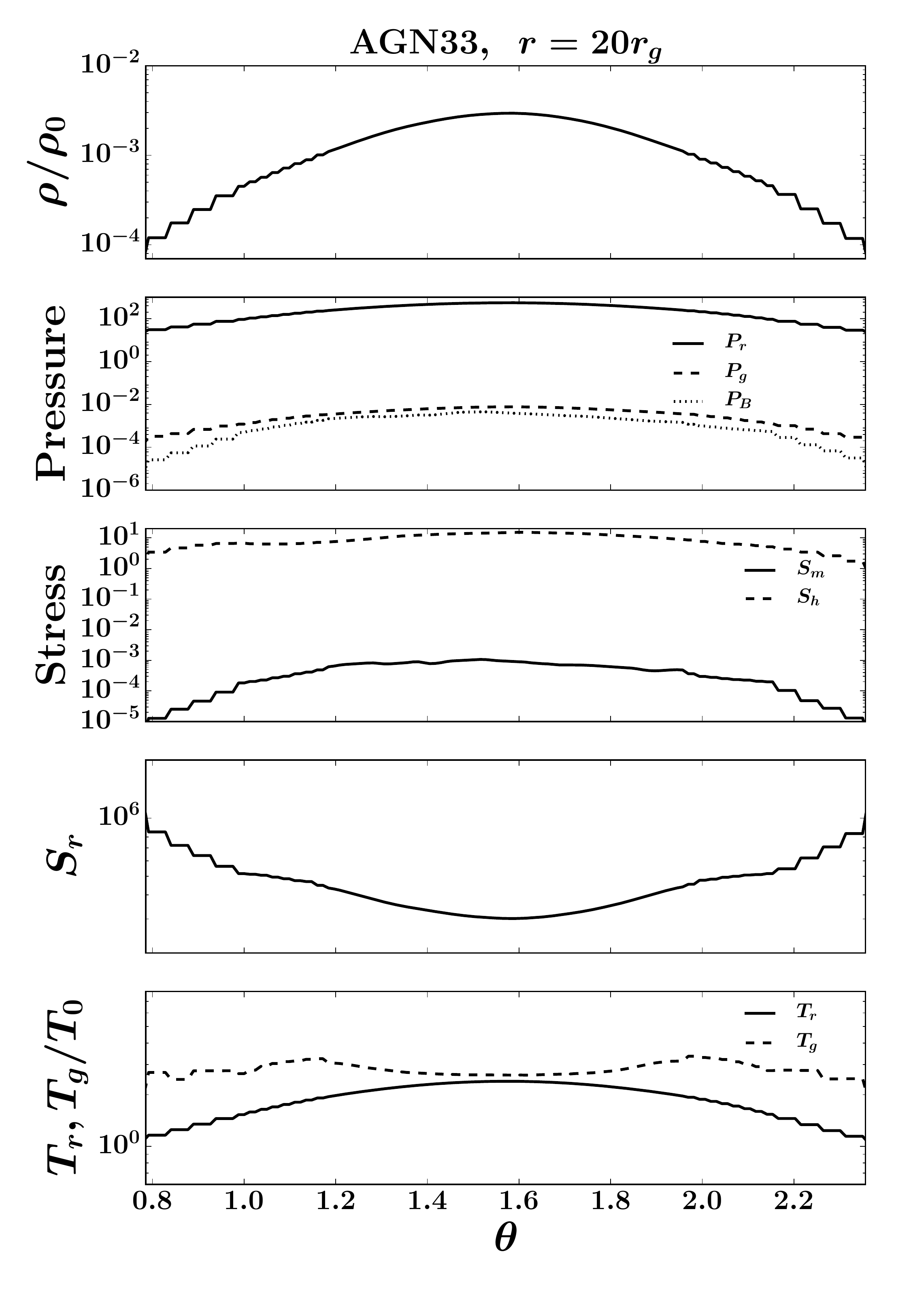}
	\includegraphics[width=0.9\hsize]{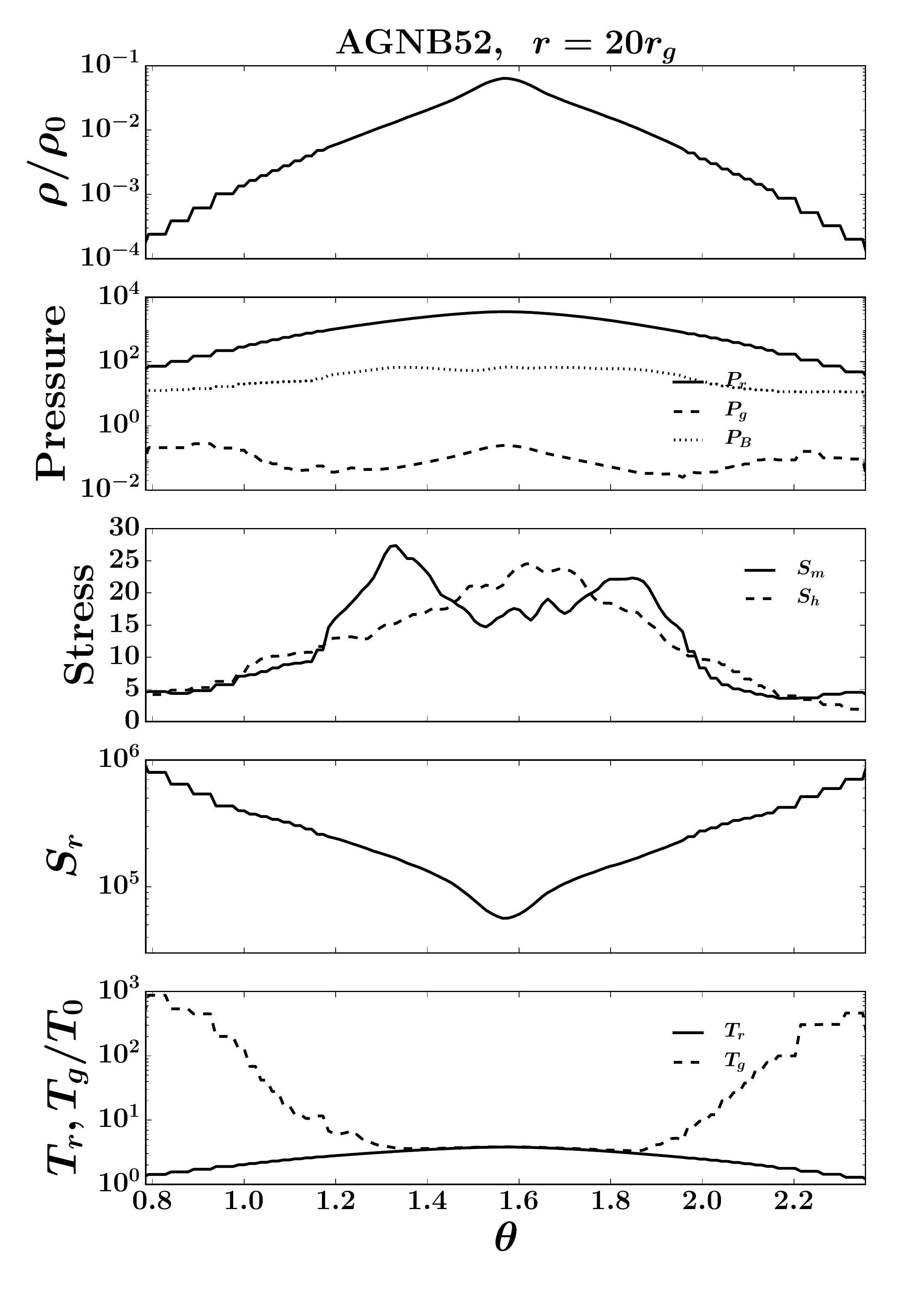}
	\caption{Time and shell averaged vertical profiles of various quantities at radius $20r_g$ for the two runs {\sf AGN33} (the top subplot) and 
		{\sf AGNB52} (the bottom subplot). From top to bottom panels of each plot, they are density $\rho$, gas ($P_g$), radiation ($P_r$) and magnetic ($P_B$) pressure, Reynolds ($S_h$) and Maxwell ($S_m$) stresses , radiation entropy ($S_r$), gas ($T_g$) and radiation ($T_r$) temperatures. 
	All the variables are scaled in the units listed in Table \ref{Table:parameters} and the unit for radiation entropy is $P_0/(\rho_0T_0)$. }
	\label{fig:vertical}
\end{figure}

\begin{figure}[htp]
	\centering
	\includegraphics[width=1.0\hsize]{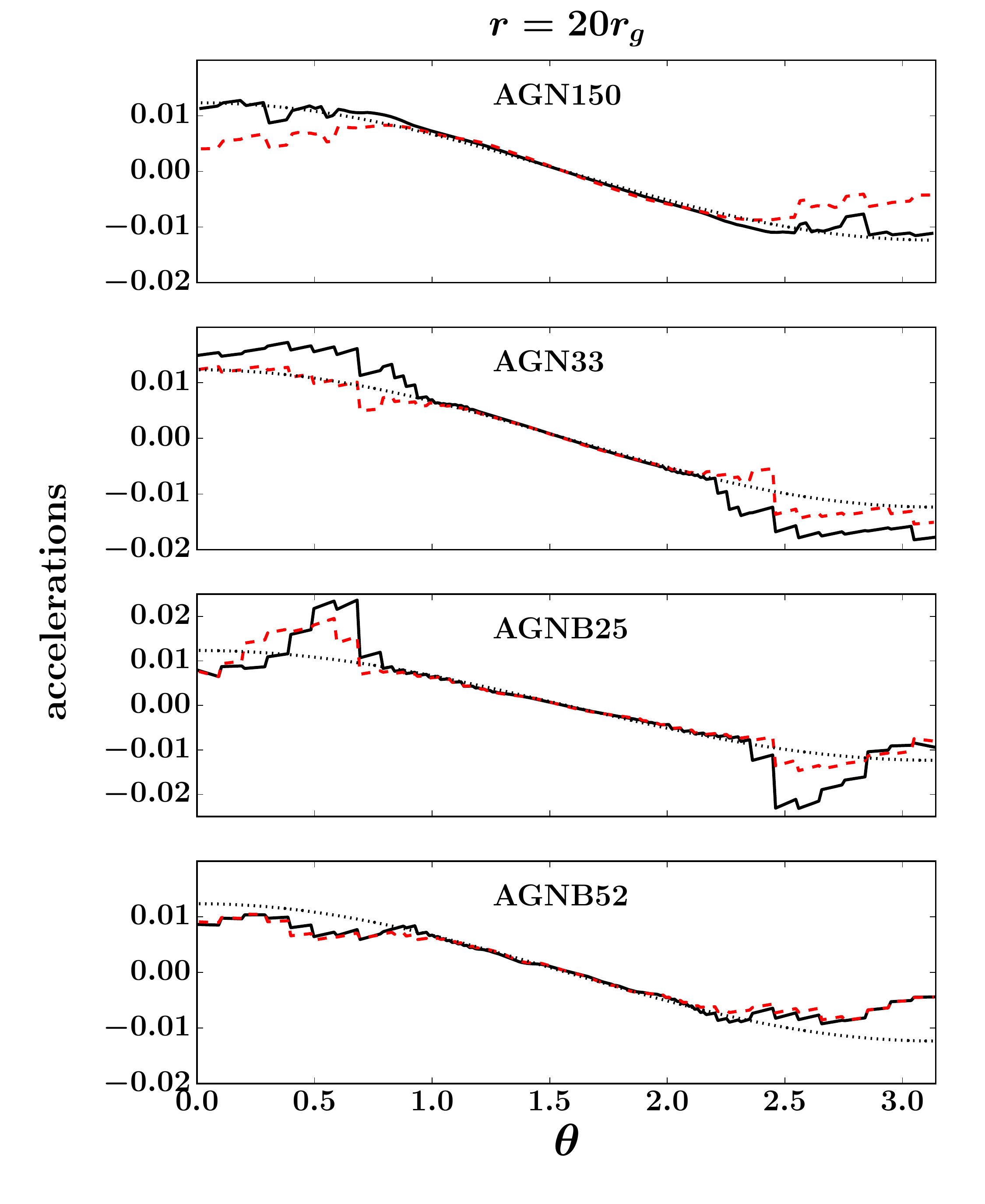}
	\caption{Time and shell averaged vertical profiles of the vertical components of radiation accelerations  at radius $20r_g$ compared with the gravitational accelerations for the four simulations. The solid black lines are the volume averaged radiation accelerations $a_r$ while the dashed red lines are the density weighted radiation accelerations (section \ref{sec:vertical}). The dotted black lines are the corresponding vertical component of gravitational accelerations. All these accelerations are scaled with $c^2/r_g$. }
	\label{fig:force}
\end{figure}

\subsection{Vertical Structures of the Disks}
\label{sec:vertical}
We use the disks at $20r_g$ from simulations {\sf AGN33} and {\sf AGNB52} as examples to study the vertical structures of the disks, 
which are shown in Figure \ref{fig:vertical}. The shapes of vertical density profiles are very similar to those produced in local shearing box simulations of radiation pressure dominated accretion disks \citep[][]{Hiroseetal2009,Jiangetal2016a}, although the density scale height can be quite different. 
Consistent with the space-time diagram shown in Figure \ref{STplot}, the density is more 
centrally concentrated in {\sf AGNB52} with a smaller density scale height than the {\sf AGN33} run, although the midplane radiation pressure is larger in {\sf AGNB52} corresponding to the higher accretion rate. Within $\sim 45^{\circ}$ from the disk midplane, density drops by a factor of $250$ with height in {\sf AGNB52} while it only drops by a factor of $24.5$ in {\sf AGN33}. 
This is probably caused by the stronger vertically advective cooling due to magnetic buoyancy associated with the MRI dynamo \citep[][]{Jiangetal2014c}. It also demonstrates that density profiles in radiation pressure dominated accretion disks are determined by the dissipation profiles, not by the hydrostatic equilibrium. 

Radiation pressure dominates in the midplane of the disk, but towards the poles magnetic pressure becomes comparable to and then larger than radiation pressure within $\sim 40^{\circ}$ from the rotation axis in the run {\sf AGNB52}. Gas temperature is in thermal equilibrium with the radiation temperature in the disk midplane and hot gas with temperature $\sim 10^8K$ is formed in the funnel region. Maxwell stress has two peaks offset from the center by $10^{\circ}$ while Reynolds stress peaks at the disk midplane as it follows the density profile. These properties are very similar to the vertical disk structures as studied by 
local shearing box simulations of MRI turbulence \citep[][]{MillerStone2000,Blaesetal2007,Jiangetal2014}. For the run {\sf AGN33} where magnetic field is not significantly amplified, radiation pressure dominates over the whole disk. Consequently, the gas temperature never increases above $6\times 10^5 K$ and there is no hot gas formed in this case (see more discussions in Section \ref{sec:obs}). Reynolds stress is larger than Maxwell stress by a factor of $\sim 10^5$ at all height. It peaks at the disk midplane and only drops by a factor of $\sim 4$ at $45^{\circ}$ from the disk midplane, which is much smaller compared with the change of density with height. Therefore, dissipation per unit mass due to the Reynolds stress is also larger in the low density regions. However, it does not produce the hot gas easily compared with the case when dissipation in the low density region is dominated by the Maxwell stress. This is probably because  dissipation caused by the Reynolds stress happen at the locations of the shocks (see Section \ref{sec:stress}), where density is relatively larger compared with the azimuthally averaged values and the energy can be converted to the photons in a short time scale. Dissipation associated with the Maxwell stress usually happens at locations where magnetic energy is larger and density is smaller than the azimuthally averaged values. Therefore, 
gas can be heated up without cooling easily. 

Vertical profiles of radiation entropy $S_r=4E_r/(3\rho T_r)$ for the two runs are shown in the fourth panels of Figure \ref{fig:vertical}, which is the dominant entropy in the strongly radiation pressure dominated regime. Because radiation energy density decreases slower with height compared with density, $S_r$ always increases with height. This suggests that these strongly radiation pressure dominated  disks are stable to the thermally driven convection for both Reynolds stress and Maxwell stress dominated cases, in contrast to the previous suggestions that convection may happen in this regime if dissipation per unit mass is assumed to be a constant \citep[][]{ShakuraSunyaev1978,Agoletal2001,ZhuNarayan2013}.

%hydrostatic equilibrium force balance, compare radiation acceleration and gravitational acceleration
Because the dynamic time scale is much shorter than the thermal time scale, we expect the vertical hydrostatic equilibrium to be maintained in the main body of the disk, which is checked at $20r_g$ for the four runs in Figure \ref{fig:force}. As radiation pressure is significantly larger than gas and magnetic pressure in the four runs, we focus on the comparison between radiation and gravitational acceleration.
The volume averaged vertical component of radiation acceleration at each $\theta$ is calculated as 
$a_r=\<\kappa_{aR}+\kappa_s\>\left(\<F_{r,r}\>\cos\theta-\<F_{r,\theta}\>\sin\theta\right)/c$, where $F_{r,r}$ and $F_{r,\theta}$ are the radial and poloidal components of radiation flux.  For turbulent flow with large density fluctuations, more photons can go through the low density regions and cause  an anti-correlation between density and radiation flux fluctuations\footnote{This is often referred to as ``porosity" in the massive star literature \citep[][]{Jiangetal2015}.} works to reduce the effective radiation acceleration. This effect can be shown by comparing $a_r$ with the density weighted radiation acceleration $a_{r,\rho}$, which is calculated as $a_{r,\rho}=\<\rho\left(\kappa_{aR}+\kappa_s\right)\left(F_{r,r}\cos\theta-F_{r,\theta}\sin\theta\right)\>/\left(c\<\rho\>\right)$. The vertical component of gravitational acceleration is $a_g=G\mbh\cos\theta/(r-2r_g)^2$. These accelerations are compared in Figure \ref{fig:force}. In the main body of the disks (within $\sim 30^{\circ}$ from the disk midplane), gravitational acceleration is balanced by the radiation acceleration, which confirms that the disk is in hydrostatic equilibrium. The density weighted radiation acceleration is also the same as the volume averaged radiation acceleration in the optically thick part of the disk for all the four runs, which is consistent with the results found in \cite{Jiangetal2015}. Above and below the disk  midplane, $a_r$ generally becomes larger than $a_{r,\rho}$ due to the density fluctuations. For the run {\sf AGN150}, although $a_r\sim a_g$ in the funnel regions, $a_{r,\rho}$ is actually smaller than $a_g$, which is consistent with the inflow gas at $20r_g$ shown in Figure \ref{averhov}. For the run {\sf AGN33} and {\sf AGNB25}, $a_{r,\rho}$ is slightly larger than the gravitational acceleration consistent with the launching of radiation driven outflow in this region (Figure \ref{averhov}). For the run {\sf AGNB52}, radiation acceleration is smaller than the gravitational acceleration in the funnel regime and magnetic fields provide additional support to gravity, because magnetic energy density is larger than the radiation energy density in this region.

\subsection{Angular Momentum Transfer Caused by the Density Waves}
\label{sec:stress}
In the previous section, we have shown that the relative contribution of Reynolds stress and Maxwell stress
to the angular momentum transfer in the disk can be quite different 
for the simulations, depending strongly on the net poloidal magnetic flux present in the flow. Contrary to all previous work on MRI driven accretion flows, the 
Reynolds stress exceeds the Maxwell stress and dominates angular momentum transfer. In this section, we will 
first show that in all situations, the simulations self-consistently conserve the angular momentum in the disk. 
Then we will show that the dominant Reynolds stress is consistent with the properties of the spiral shocks formed in the simulations. 

\subsubsection{The Angular Momentum Flux}
Evolution of the sum of gas and radiation momentum is determined by the following 
equation 
\begin{eqnarray}
\frac{\partial}{\partial t}\left(\rho\bv+\frac{\bF_r}{c^2}\right)+\bfnabla\cdot\left(\rho\bv\bv-\bb\bb+{\sf P}^{\ast}+{\sf P_r}\right)=\nonumber\\
-\rho\bfnabla\phi.
\end{eqnarray}
We multiply the above equation with the radial vector $\br$ and get the evolution 
equation for the  $z$ component of the angular momentum due to the radial angular momentum transport as 
\begin{eqnarray}
\frac{\partial r\sin\theta\rho v_{\phi}}{\partial t}+\frac{1}{r^2}\frac{\partial }{\partial r}\left(r^2F_J\right)=0.
\label{eqn:ang}
\end{eqnarray}
Here $F_J$ is the radial angular momentum flux
\begin{eqnarray}
F_J\equiv r\sin\theta\left(\rho v_rv_{\phi}-B_rB_{\phi}+P_{r,r\phi}\right),
\label{eqn:FJ}
\end{eqnarray}
where $P_{r,r\phi}$ is the $r-\phi$ component of the radiation pressure tensor. Here we only consider 
angular momentum flux along radial direction. Because gravity is provided by the 
central point source in the simulations, only the local Reynolds, Maxwell and radiation stresses will 
provide angular momentum transfer at each radius. We also neglect the term $\bF_r/c^2$ 
as radiation momentum is usually much smaller than the momentum of the gas.

\begin{figure}[htp]
	\centering
	\includegraphics[width=1.0\hsize]{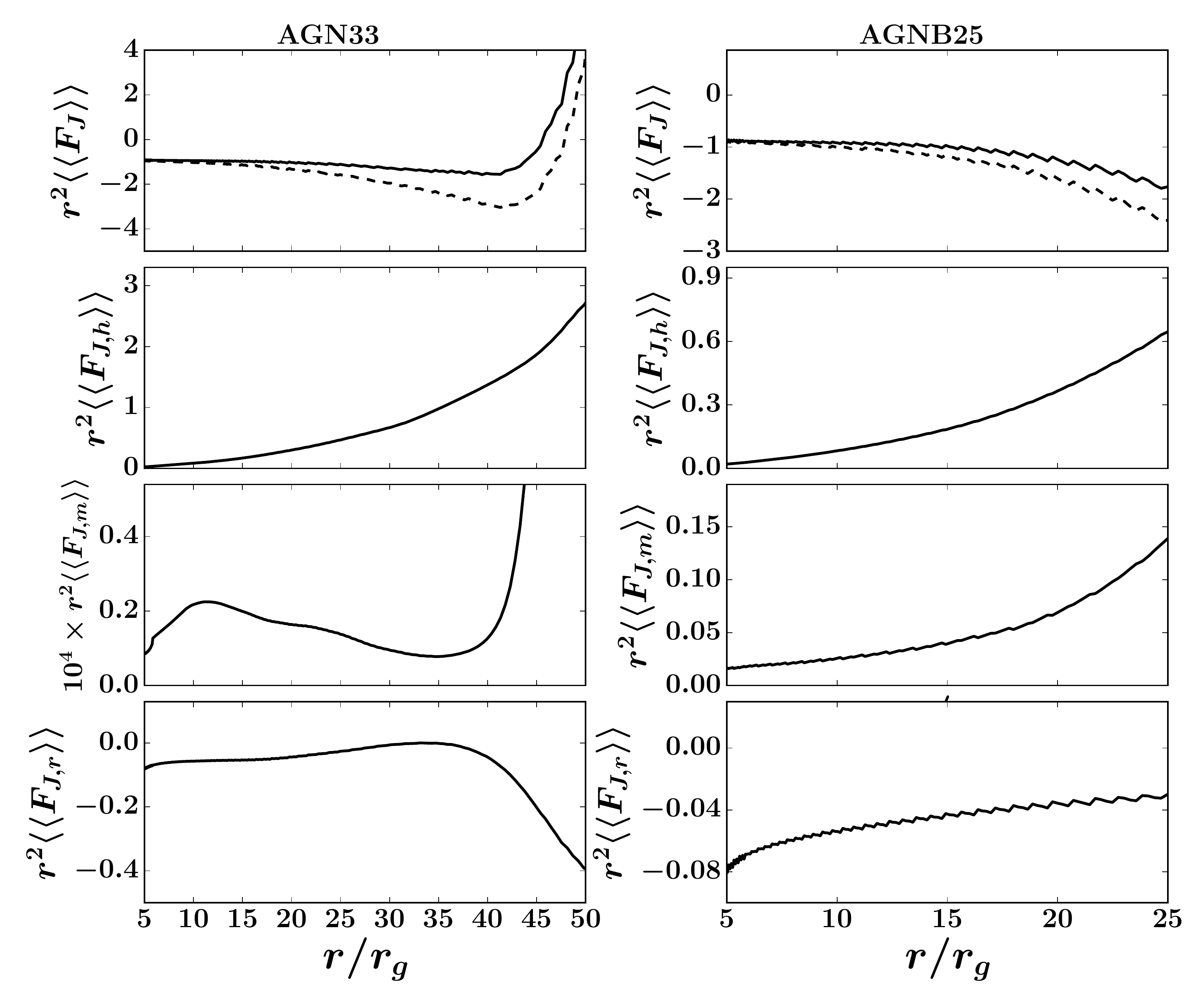}
	\caption{Radial profiles of the time and azimuthally averaged total angular momentum flux $r^2F_J$ (solid lines in the top panel), angular momentum flux due to Reynolds stress $r^2F_{J,h}$ (the second panels), Maxwell stress $r^2F_{J,m}$ (the third panels) 
		and radiation stress (the fourth panel). The dashed lines in the top panels 
		are the mean angular momentum flux carried by the net mass flux $r^2\<\rho v_r\>\<v_{\phi}\>$. The left column is for the run 
		{\sf AGN33} while the right column is for the run {\sf AGNB25}. Units of the angular momentum flux $F_J,F_{J,h},F_{J,m}$ are all 
	$r_gP_0$. }
	\label{ang_flx}
\end{figure}

The hydrodynamic stress can be decomposed to the mean and fluctuation components
\begin{eqnarray}
\rho v_r v_{\phi}=(\delta \rho v_r + \< \rho v_r\>)(\delta v_{\phi}+\<v_{\phi}\>),
\end{eqnarray}
where the fluctuation quantities are defined as $\delta \rho v_r\equiv \rho v_r-\<\rho v_r\>$, 
$\delta v_{\phi}=v_{\phi}-\<v_{\phi}\>$. The mean hydrodynamic stress is then the sum of the 
mean Reynolds stress and the angular momentum carried by the mean accretion flow
\begin{eqnarray}
\<\rho v_r v_{\phi}\>&=&\<\delta \rho v_r\delta v_{\phi}\>+\<\rho v_r\>\<v_{\phi}\>.
\end{eqnarray}

In steady state, conservation of angular momentum requires that the time and azimuthally averaged 
total angular momentum flux $r^2\<F_J\>$ is independent of radius $r$, which is checked in Figure \ref{ang_flx} 
for the two runs {\sf AGN33} and {\sf AGNB25}. The contributions of angular momentum 
flux by the  Reynolds stress $F_{J,h}\equiv r\sin\theta \delta \rho v_r\delta v_{\phi}$, Maxwell stress 
$F_{J,m}\equiv -r\sin\theta B_rB_{\phi}$ and radiation stress $F_{J,r}=r\sin\theta P_{r,r\phi}$ are also shown in this Figure. 
Inside $\sim 40r_g$ for the run {\sf AGN33} 
and $\sim 20r_g$ for {\sf AGNB25}, $r^2\<F_J\>$ is indeed roughly a constant. Deviations in larger radii 
are because the simulations are not run long enough to reach steady state there. In run {\sf AGN33}, 
angular momentum transport is dominated by the Reynolds stress, while Maxwell stress is completely negligible. 
As shown in Section \ref{sec:radial}, the radiation contribution $F_{J,r}$ is dominated by the drag of photons due to the mean rotation and inflow 
of the gas in the disk. The situation is very similar for the run {\sf AGN150}. 
For run {\sf AGNB25} with net poloidal magnetic fields, angular momentum flux caused by the Maxwell stress is significantly increased and it is only 
a factor of $\sim 4$ smaller than the Reynolds stress. When the net poloidal magnetic flux is significantly increased in {\sf AGNB52}, the Maxwell stress becomes the dominant angular momentum transfer mechanism.

\begin{figure}[htp]
	\centering
	\includegraphics[width=1.0\hsize]{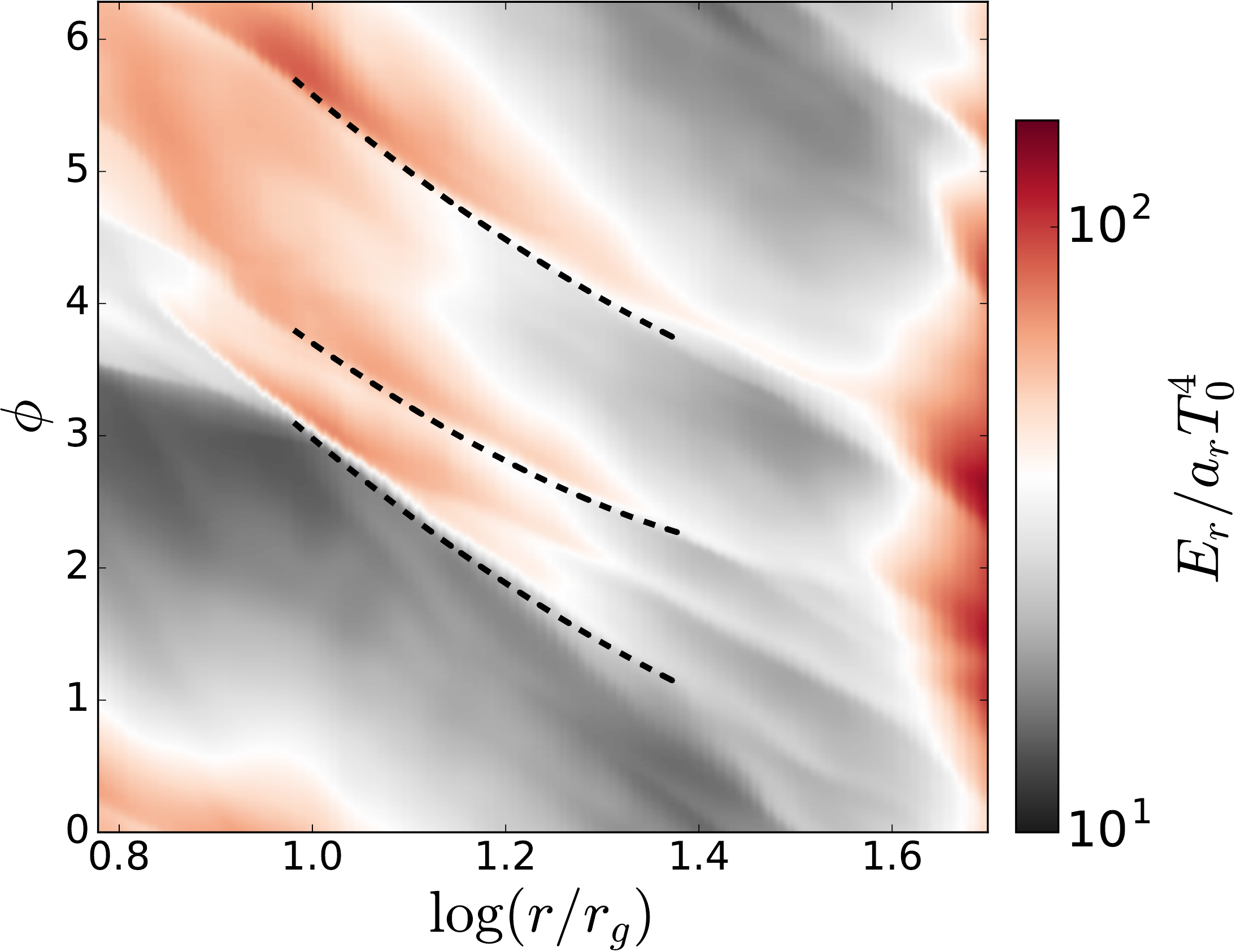}
	\caption{Radial and azimuthal distribution of the volume averaged radiation energy density at the time $2.77\times 10^4r_g/c$ for the simulation {\sf AGN33}. The volume average is only done within $30^{\circ}$ above and below the disk midplane. The three dashed black lines are linear dispersion relations for density waves with pattern speed $\Omega_p=2.36\times 10^{-3}c/r_g$ and wave numbers $m=2,1.5$ and $2$ respectively (from top to bottom), which fit the locations of shocks very well.}
	\label{fitshock}
\end{figure}

\subsubsection{Properties of the Density Waves}
\label{sec:spiral}
To confirm that the spiral structure we see in our simulations (Figure \ref{spiralimage}) are indeed shocks caused by the density waves, shapes of the spiral structures can be compared with predictions of linear dispersion relation of density waves. Although shocks are clearly nonlinear phenomenon, linear theory can usually do a good job to predict the locations of the shocks (\citealt{Juetal2016}).

To determine the pattern speed of the density waves, at each snapshot, we first take slices of radiation energy density $E_r$ through the disk midplane. Then we calculate the discrete spatial fourier transform of $E_r$ over $\phi$ direction for each radius $r$. The spatial fourier power spectra do not show a single value of spatial wavenumber $m$. Instead, the spectra are broadly distributed between $m=1$ and $m=10$ with peaks around $m=2$ and $3$. The spectra drops significantly with increasing wavenumber when $m>10$. There are more power with larger wavenumber at small radii.  For $r>20r_g$, most of the power is concentrated within $m<5$. 

We then take the temporal fourier transform over the snapshots of the real part of the spatial fourier transfer for the $m=1$ mode at each radius. The time intervals for the temporal fourier transfer are chosen to be the same as we use for the time average as described in Section \ref{sec:spatial}. Each radius also shows multiple peaks in the temporal fourier transform. However, there is a common peak across different radii, which corresponds to the patten frequency of the density wave. With this approach, we find the pattern speeds $\Omega_p$ for the four simulations 
${\sf AGN150,AGN33,AGNB25,AGNB52}$ are $ 1.03\times 10^{-3}, 2.36\times10^{-3},2.28\times 10^{-3},1.44\times10^{-3} c/r_g$ respectively. These pattern speeds equal Keplerian rotation speeds at $99.4,57.8,59.1$ and $79.8r_g$. With these pattern speeds, for each spatial wavenumber $m$, the linear dispersion relation for density waves describing the relationship between radius $r$ and azimuthal angle $\phi$ are \citep[][]{BinneyTremaine2008,Juetal2016}
\begin{eqnarray}
d\phi=-\frac{k_r}{m}dr=-\frac{1}{c_{r,s}}\sqrt{\left(\Omega-\Omega_p\right)^2-\kappa^2/m^2}dr. 
\label{eqn:linear}
\end{eqnarray} 
Here $k_r$ is the radial wavenumber, $\kappa$ is the epicycle frequency and $c_{r,s}\equiv \sqrt{4P_r/\left(3\rho\right)}$ is the adiabatic radiation sound speed. We have picked the trailing branch of the dispersion relation corresponding to what we see in the simulations. We use the azimuthally averaged $c_{r,s}, \Omega\equiv v_{\phi}/r$ at each radius in the dispersion relation and $\kappa$ is taken to be the same as $\Omega$. A snapshot of radiation energy density $E_r$ at time $2.77\times 10^4r_g/c$ averaged within $30^{\circ}$ from the disk midplane for simulation {\sf AGN33} is shown in Figure \ref{fitshock}. There are clearly three shocks between $8r_g$ and $31.6r_g$. We pick three starting points and integrate the dispersion relation to determine the shape of the spiral structures, which are shown as the dashed black lines in Figure \ref{fitshock}. If we choose the wavenumber 
$m$ to be $2,1.5$ and $2$ for the three arms, the dispersion relation can fit the spiral shock positions very well. Notice that we need to
choose a non-integer $m$ value to describe the middle shock, which means that shock is probably formed by merger of multiple linear modes.  
At each radius along the azimuthal direction, when we cross each shock, density at the shock downstream increases while $v_{\phi}$ drops. At the same time, radial inflow velocity develops. This is how the shock driven accretion works. The shock causes the fluid element to lose angular momentum and  
flow inward. Higher radiation energy density at the shock downstream also means these are the locations where dissipation occurs.

\begin{figure}[htp]
	\centering
	\includegraphics[width=1.0\hsize]{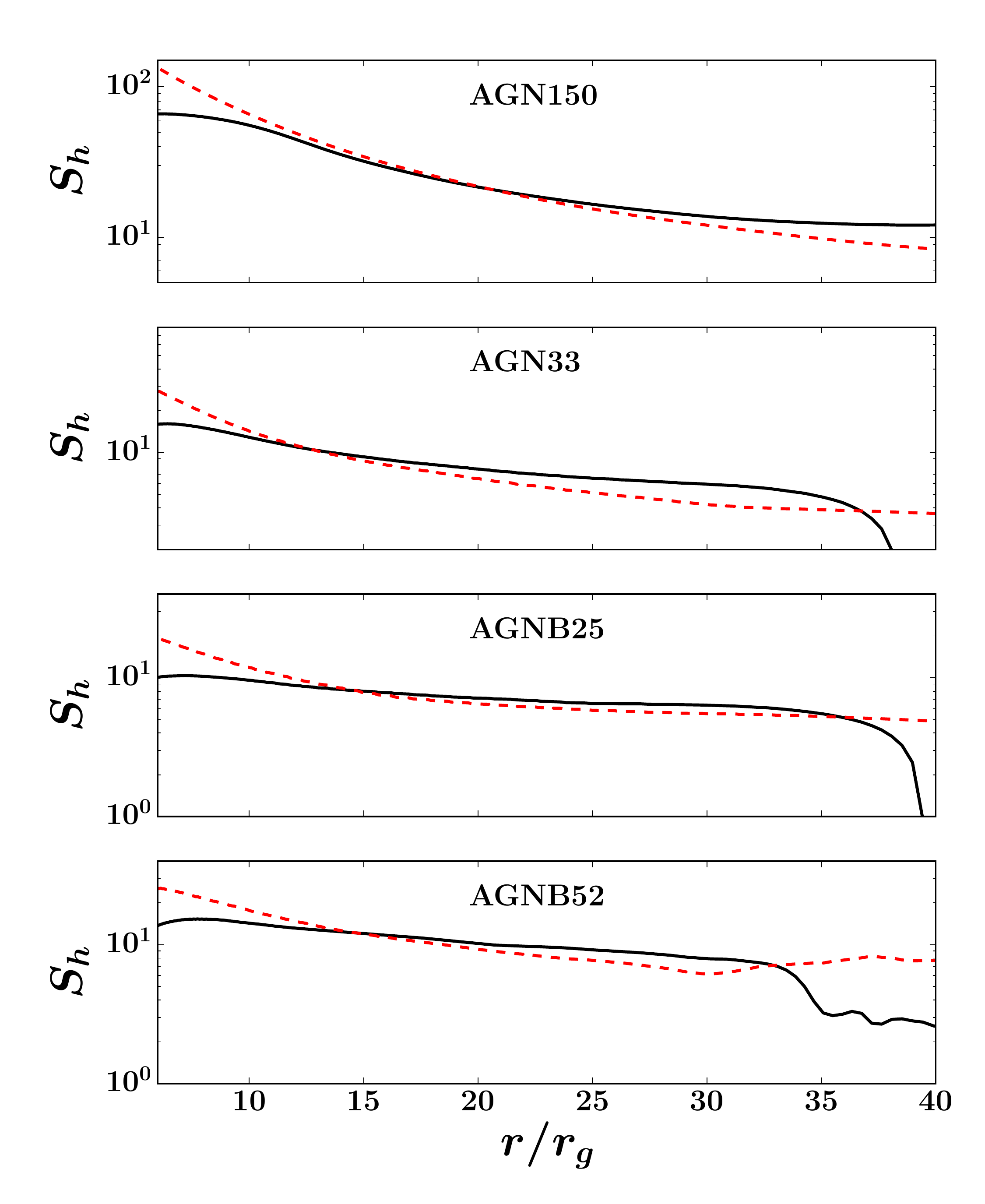}
	\caption{Compare the Reynolds stress calculated in the simulations with the predicted values from linear dispersion relations of density waves (equation \ref{eqn:fitstress}). The dashed red lines are the radial profiles of the time and volume averaged Reynolds stress while the solid black lines are the predicted values with $m=2$. The Reynolds stress we get is comparable to the predicted value. }
	\label{fitstress}
\end{figure}

\subsubsection{Reynolds Stress due to the Density Waves}
%In MRI turbulence studied by local shearing box simulations \citep[][]{Stoneetal1996,Turneretal2003,Sanoetal2004,Davisetal2010} as well as global %simulations not in the radiation pressure dominated regimes %\citep[][]{Armitageetal2001,HawleyKrolik2001,Sorathiaetal2012,Hawleyetal2011,Hawleyetal2013}, 
Based on the linear analysis of density waves, the amount of Reynolds stress generated by the density waves is proportional to the turbulent radial kinetic energy as \citep[][Equation 43]{Balbus2003}
\begin{eqnarray}
\<\delta\rho v_r\delta v_{\phi}\>&=&\frac{m}{rk_r}\left(1-\frac{\kappa^2}{m^2\left(\Omega-\Omega_p\right)^2}\right) \<\rho\delta v_r^2\>\nonumber\\
&=&\frac{c_{r,s}}{r}\frac{\sqrt{\left(\Omega-\Omega_p\right)^2-\kappa^2/m^2}}{\left(\Omega-\Omega_p\right)^2}\<\rho\delta v_r^2\>.\nonumber\\
\label{eqn:fitstress}
\end{eqnarray}
Here we calculate the turbulent radial kinetic energy as $ \<\rho\delta v_r^2\>=\<\rho v_r^2\>-\<\rho\>\<v_r\>_{\rho}^2$ and the linear dispersion relation (equation \ref{eqn:linear}) is used to calculate $m/(rk_r)$. As multiple modes of density waves are excited in the simulations, we just pick a value of $m$ to compare the Reynolds stress calculated from the simulations and the predicted stress based on this formula with the pattern speed $\Omega_p$ we have identified in the last section. The results with $m=2$ for the four simulations are shown in Figure \ref{fitstress}, which shows that the Reynolds stress given by the linear analysis of density waves is pretty close to what we get from the simulations. The Reynolds stress from $m=2$ mode always under-predicts the stress in the inner region ($r\lessapprox 15r_g$) and over-predicts the stress in the outer region ($r\gtrapprox 15r_g$). This is consistent with what we find in the last section that there are more high wavenumber modes in the small radii and more low wavenumber modes in the large radii. 

%\begin{figure}[htp]
%	\centering
%	\includegraphics[width=1.0\hsize]{LC_history.pdf}
%	\caption{History of total  emerging radiation luminosity $L_r$ (red lines) and kinetic energy flux $L_k$ associated with the outflow (black lines) inside %$40r_g$ measured at the outer boundary of the simulation box. The top panel is for the run {\sf AGNM2} while the bottom panel is for the run {\sf %AGNBM1}.}
%	\label{lc}
%\end{figure}

\subsection{Properties of the Emerging Photons and Outflow}
\label{sec:outflow}
\begin{figure*}[htp]
	\centering
	\includegraphics[width=0.48\hsize]{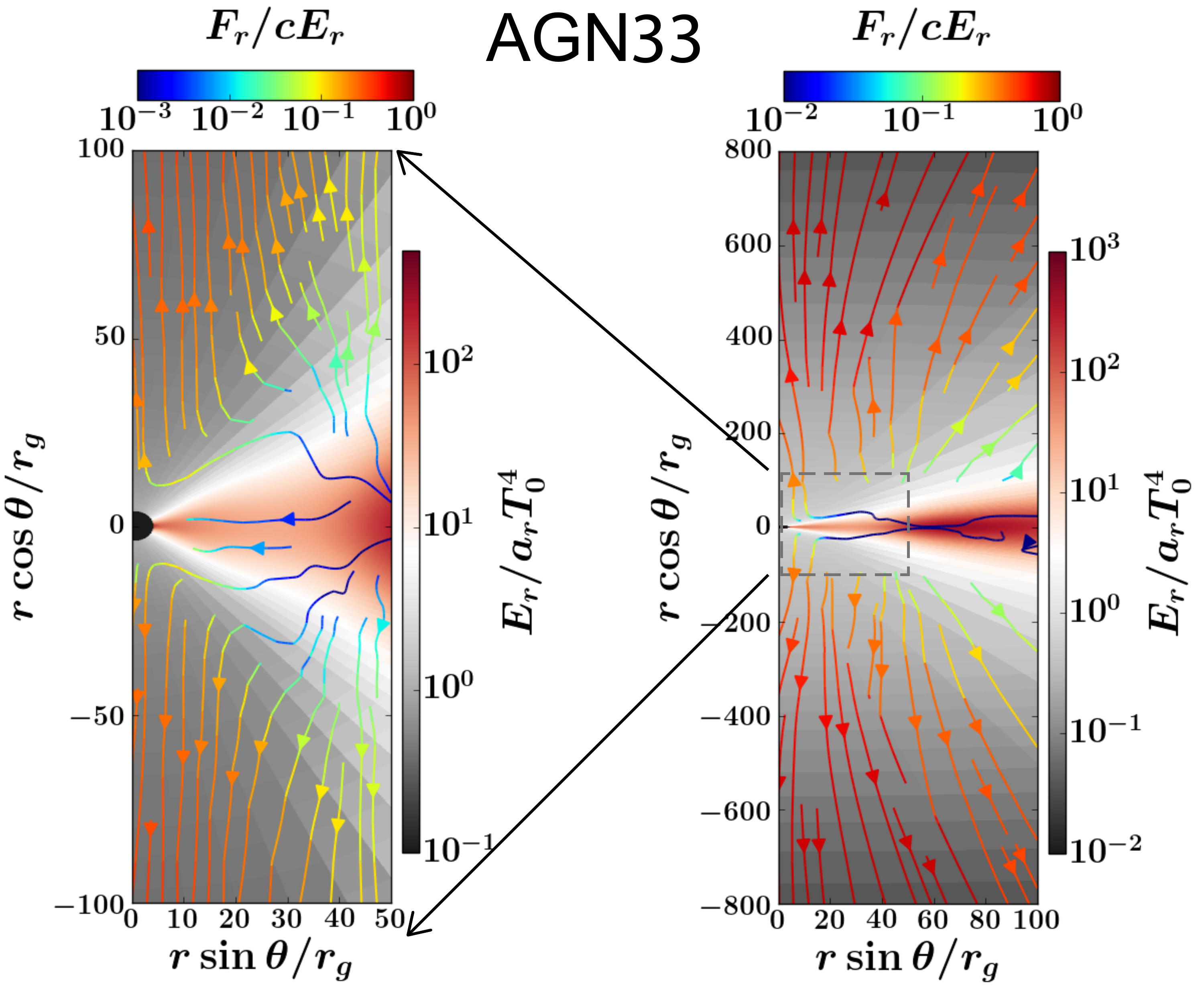}
    \includegraphics[width=0.48\hsize]{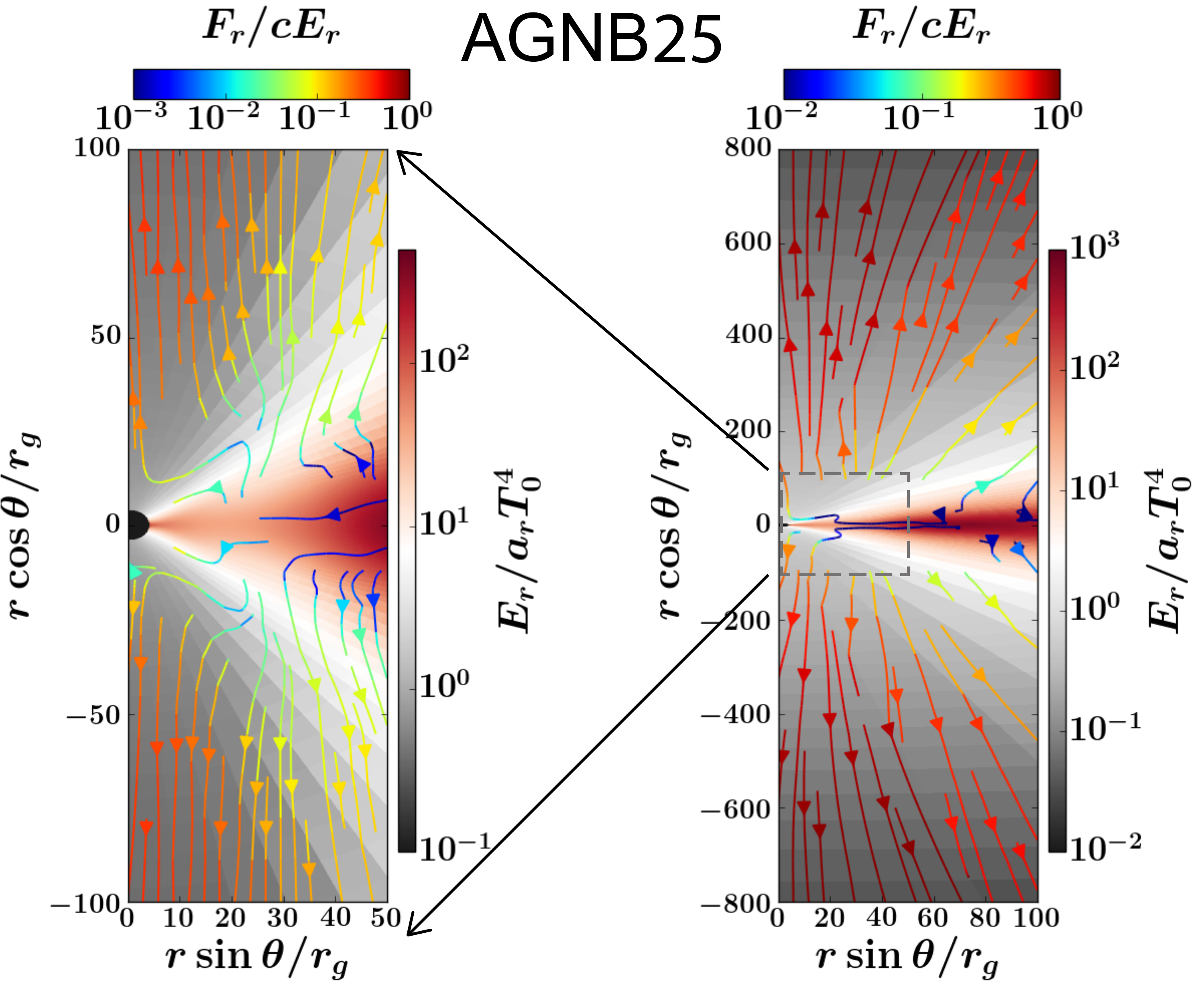}
	\caption{Time and azimuthally averaged spatial structures of radiation energy 
		density $E_r$ (colors) and radiation flux $\bF_r$ (streamlines, only includes the azimuthally averaged 
		$r,\theta$ components $F_{r,r},\ F_{r,\theta}$). 
        Color of the streamlines is the ratio  $|\bF_r|/\left(cE_r\right)$. 
        The left two panels are for the run {\sf AGN33} while the right two panels are for the run
        {\sf AGNB25}. In each case, the first panel shows the inner structures with height smaller than 
        $100r_g$ while the second panel show the disk structures up to $800r_g$ in height. }
	\label{Frwind}
\end{figure*}

Outflow coming from the funnel region of the disk along the rotation axis is a natural outcome of the accretion disks with accretion rates above the Eddington limit. That is also the region where most of the photons escape. 
Shape of the funnel region can be quantified by the location where total optical depth measured from the rotation axis $\tau_R$ reaches one, which is calculated as 
\begin{eqnarray}
\tau_R=\int_0^R\left(\kappa_s+\kappa_{aR}\right)\rho dR.
\end{eqnarray}
This is shown as the dashed black lines in Figure \ref{averhov}. Because of the mesh structure and logarithmic grid we use, the funnel is better resolved near the disk midplane and poorly resolved at larger radius. For simulation {\sf AGN150} with $\dot{M}=150M_{\rm Edd}$, almost all the funnel region becomes optically thick. Particularly, the outflow only develops from $\sim 50r_g$. For the other runs, the outflow starts $\sim10r_g$.
The optically thin part of the funnel is also small in {\sf AGNB52} because density drops very slowly with height due to strong magnetic pressure and the net accretion rate is also larger than the values in {\sf AGN33} and {\sf AGN150}. 
Radiation energy density and radiation flux streamlines ($F_{r,r}$ and $F_{r,\theta}$ components only) for the two runs {\sf AGN33} and {\sf AGNB25} are shown in Figure \ref{Frwind}, which also shows the ratio between radiation flux magnitude and $cE_r$. This figure shows the large scale structures of the radiation flux within $800r_g$ in height as well as the inner region of the disks within $100r_g$. 
The ratio $|\bF_r|/\left(cE_r\right)$ is almost zero near the optically thick midplane and reaches $\sim0.4$ at $r\sim 200r_g$. Most of the photons leaving in the outflow region can trace back to the inner $\sim 40r_g$ with streamlines nearly parallel to the rotation axis. When $r\gtrsim 200r_g$, the angle between rotation axis and radiation flux increases as density from the disk drops and it  is not optically thick enough to collimate the photons.

\begin{figure}[htp]
	\centering
	\includegraphics[width=0.9\hsize]{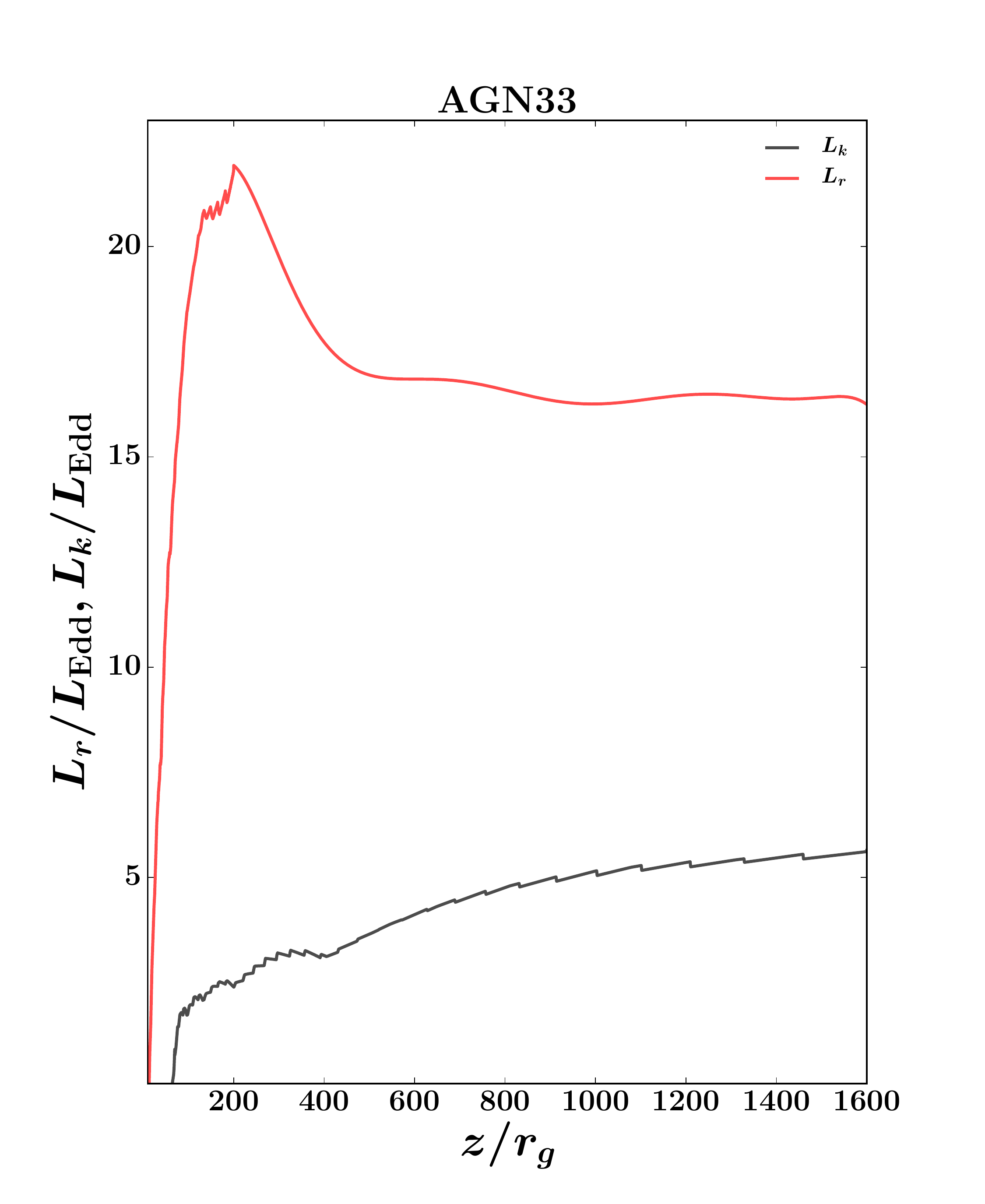}
	\includegraphics[width=0.9\hsize]{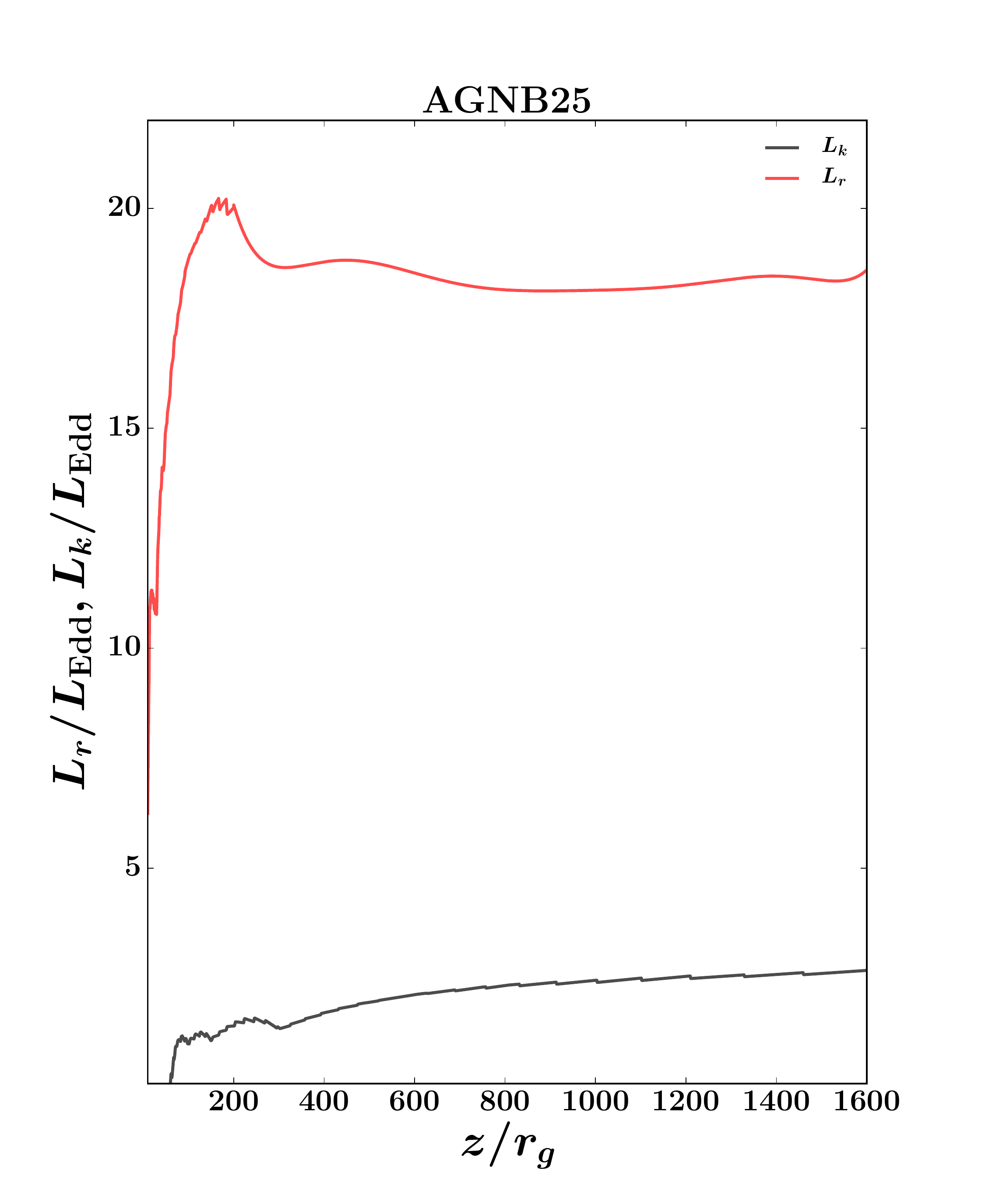}
	\caption{Vertical profiles of total radiative luminosity ($L_r$, red lines) and kinetic luminosity associated with the unbound outflow ($L_k$, black lines) for the simulations {\sf AGN33} (top panel) and {\sf AGNB25} (bottom panel). The luminosities are calculated through the surface of 
   cylinders with radii $R=40r_g$ and $R=30r_g$ for the two runs {\sf AGN33} and {\sf AGNB25} respectively.  Both $L_r$ and $L_k$ reach roughly constant values with height near the boundary of our simulation domain and $L_r$ is significantly larger than $L_k$ in the simulations. We have smoothed the curves of $L_r$ to remove the oscillations due to the coarse grid at large radii. }
	\label{fig:lumprofile}
\end{figure}

%\begin{figure}[htp]
%	\centering
%	\includegraphics[width=1.0\hsize]{Frwind.pdf}
%	\caption{Time and azimuthally averaged spatial structures of radiation 
%		flux $\bF_r$ (streamlines) and the ratio $|\bF_r|/\left(cE_r\right)$ (color) 
%		for the two runs {\sf AGNM2} and {\sf AGNBM1} respectively. 
%        Radiation flux $\bF_r$ only includes the azimuthally averaged 
%        $r,\theta$ components $F_{r,r},\ F_{r,\theta}$.  }
%	\label{Frwind}
%\end{figure}

To calculate the total radiative and kinetic energy luminosity carried with the outflow, we use the fluxes through a cylindrical surface of radius $R_0 \sim 30-40 r_g$.  We use this surface rather than the radial boundary of our simulation domain because the radiation and outflow through this surface is driven primarily by the accretion in the inner 30-40 $r_g$, where our simulations have reached a steady state.  If we extended this surface further out, it would start to encompass regions where the emission and outflow are dominated by the still relaxing portions of our initial torus. We first convert the azimuthally averaged flow velocity $\bv$, mass flux $\rho\bv$ and radiation flux $\bF_r$ from $r,\theta$ plane in the spherical polar coordinate system to the $R,z$ plane in the cylindrical coordinate. Then the total radiative luminosity $L_r$, kinetic energy luminosity $L_k$ and mass flux 
$\dot{M}_w$ leaving from a fixed height $z$ inside the cylindrical radius $R_o$ are calculated as
\begin{eqnarray}
L_r&=&\int_0^{R_o} 2\pi F_{r,z}RdR+\int_{-z}^{z} 2\pi R_0 F_{r,R}dz,\nonumber\\
L_k&=&\int_0^{R_o} 2\pi v_{z}\left(\frac{1}{2}\rho v^2\right)RdR\nonumber\\
    &+&\int_{-z}^{z} 2\pi v_{R}\left(\frac{1}{2}\rho v^2\right)R_0dz,\nonumber\\
\dot{M}_w&=&\int_0^{R_o} 2\pi \rho v_zRdR+\int_{-z}^{z} 2\pi \rho v_R R_0dz.
\end{eqnarray}
We sum contributions from both sides of the disk. For the integral along the cylindrical surface at the fixed radius $R_0$, we only 
include regions with positive $F_{r,R}$ and $v_R$ to exclude the inflow part in the main body of the disk. For $L_k$ and $\dot{M}_w$, 
we only include fluid elements when the sum of kinetic and gravitational potential energies are positive, which indicates that the gas is truly unbound. The time averaged values of $L_r,L_k$ are calculated during the time period as specified in Section \ref{sec:spatial}. 

%Histories of $L_r$ and $L_k$ for simulations {\sf AGNM2} and {\sf AGNBM1} calculated with $R_o=40r_g$ at $z=200r_g$ are shown in Figure \ref{lc}. %Compared with the history of mass accretion rate at $10r_g$ as shown in Figure \ref{mdot_hist}, the escaping photons show a much smoother light %curve as expected. They only follow the change of $\dot{M}$ on the time scale of $\sim 3000t_0$, which roughly corresponds to the thermal time scale %at $R_o$. 
%After the initial $10^4t_0$, $L_r$ varies between $4L_{\rm Edd}$ and $6L_{\rm Edd}$ while $L_k\sim L_{\rm Edd}$ for both runs, which shows that most %of the energy still leaves the disk as photons. 
Vertical profiles of $L_r$ and $L_k$ for the two runs {\sf AGN33} and {\sf AGNB25} are shown in Figure \ref{fig:lumprofile}. We choose $R_0=40r_g$ for {\sf AGN33} and $R_0=30r_g$ for {\sf AGNB25} to match the steady state part of the disk in the simulations. The radiative and kinetic luminosities increase rapidly inside $100r_g$ and $L_r$ reaches a roughly constant value beyond $z\approx 400r_g$, which is 
$16.6L_{\rm Edd}$ for {\sf AGN33} and $18.4L_{\rm Edd}$ for {\sf AGNB25} respectively.  The kinetic luminosity $L_k$ also saturates to $5.6L_{\rm Edd}$ for {\sf AGN33} 
and $2.7L_{\rm Edd}$ for {\sf AGNB25} near the boundary of our simulation domain. This demonstrates that our simulation domain is large enough to capture the energy exchange between the photons and gas. The kinetic luminosity is only $34\%$ and $15\%$ of the radiative luminosity in the two simulations. As the net mass accretion rate is $33M_{\rm Edd}$ inside $40r_g$ 
for {\sf AGN33} and $25M_{\rm Edd}$ inside $30r_g$ for the two runs respectively, the radiative efficiencies are $5.0\%$ and $7.3\%$, which are comparable to the value we get in \cite{Jiangetal2014c}.

Although Figure \ref{avemdot} shows that there is a significant fraction of gas moving outward in the disk, not all of them are  unbound outflow. At the outer boundary of our simulation domain, mass fluxes associated with the gas which has kinetic energy larger than the gravitational energy are $18.1M_{\rm Edd}$ for {\sf AGN33} and $8.1M_{\rm Edd}$ for {\sf AGNB25}, which are $54.8\%$ and 
$32.4\%$ of the net mass accretion rates. 
 
 For the run {\sf AGN150}, outflow only starts around $50r_g$, inside which all the photons are advected towards the black hole. The outflow velocity at the outer edge of the simulation box is smaller by a factor of $\sim 2$ compared with the corresponding values in {\sf AGN33} and {\sf AGNB25}. Although the outer edge of the simulation box is still very optically thick, $L_r$ shows similar properties as in Figure \ref{fig:lumprofile}, which reaches a constant value beyond $z=1200r_g$. This can happen in the optically thick region because most of the luminosity comes from the advection of the photons while radiative acceleration associated with the co-moving frame flux roughly balances the gravitational acceleration. Therefore the gas is not accelerated and radiation energy is not converted to the gas kinetic energy. At the boundary of the simulation domain, $L_r=15.5\Ledd$ while $L_k=3.6\Ledd$. As the net mass accretion rate is $150\Medd$, the radiative efficiency is only $1\%$. The mass flux associated with the unbound outflow is $23\Medd$, which is $15\%$ of the net mass accretion rate.

 Because of the limited radial range we can reach steady state for the run {\sf AGNB52}, we will not try to estimate the radiative efficiency for this run, as we will likely miss a lot of radiation flux at larger radii.  
 Although the optical depth in the funnel region is larger in this run compared with {\sf AGN33} and {\sf AGNB25}, outflow starts around $20r_g$, which is much smaller compared with the run {\sf AGN150}. Therefore, we expect the radiative efficiency in {\sf AGNB52} to be larger than the value in {\sf AGN150}, but smaller than the values in {\sf AGN33} and {\sf AGNB25}. 

\begin{figure*}[htp]
	\centering
	\includegraphics[width=1.0\hsize]{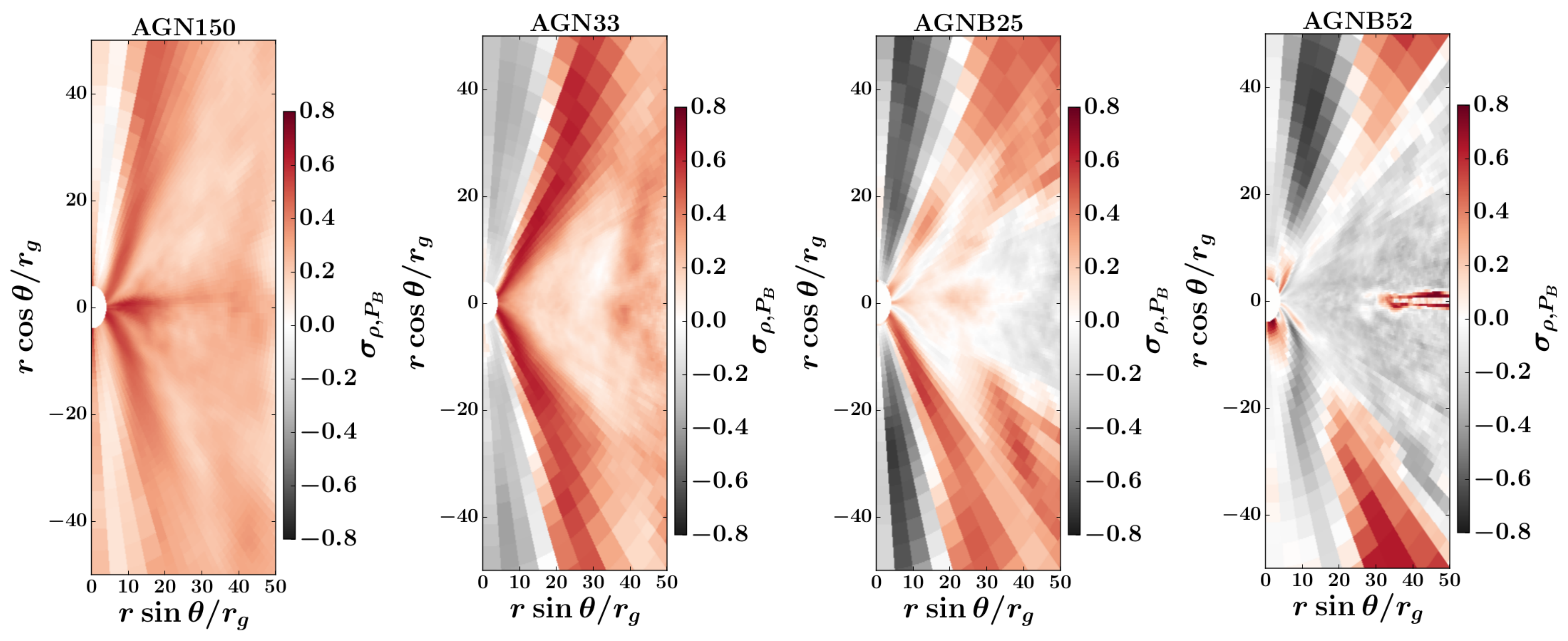}
	\caption{Time averaged cross correlation coefficients $\sigma_{\rho,P_B}$ between azimuthal density and magnetic pressure fluctuations for the four runs in the inner regions of the disks. Positive correlations are found for {\sf AGN150} and {\sf AGN33} where magnetic pressure is much smaller than the radiation pressure. In {\sf AGNB25} and {\sf AGNB52} where Maxwell stress provides a significant fraction of the total stress, negaive cross correlations between $\rho$ and $P_B$ show up.}
	\label{rhoBcorr}
\end{figure*}

\subsection{Energy Transport Mechanisms in the Disk}
The energy dissipated in the main body of the disk needs to be transported to the funnel region efficiently before the photons are advected to the black hole in order to achieve relatively large radiative efficiency compared to the values predicted by the slim disk model. \cite{Jiangetal2014c} shows that  magnetic buoyancy allows photons to move vertically with a speed (typically the local Alfv\'en speed \citep[e.g.,][]{StellarRosner1984,Jiangetal2017}) much larger than the photon diffusion speed in the super-Eddington accretion regime. The signature for this mechanism to be working is the strong anti-correlation between the azimuthal fluctuations of density and magnetic energy density \citep[][]{Blaesetal2011,Jiangetal2014c}. Time averaged cross correlation coefficients $\sigma_{\rho,P_B}$ between $\rho$ and $P_B$ for the four runs are shown in Figure \ref{rhoBcorr}. Density and magnetic pressure fluctuations are actually correlated in {\sf AGN150} and {\sf AGN33}. They become anti-correlated in the main body of the disks in {\sf AGNB25} and {\sf AGNB52}. This is because in the two runs {\sf AGN150} and {\sf AGN33} when there are no net vertical magnetic fields, Maxwell stress is weak and magnetic pressure is smaller than the radiation pressure by $\sim 10^3-10^4$ (Figure \ref{fig:radial}). While in the two runs {\sf AGNB25} and {\sf AGNB52} with net vertical magnetic fields, magnetic fields are significantly amplified and magnetic pressure can reach $\sim 10\%$ of radiation pressure. In order for the magnetic buoyancy to work, azimuthal fluctuations of magnetic pressure in the low density region should be able to balance fluctuations of gas and radiation pressure in the high density regions, which cannot happen for {\sf AGN150} and {\sf AGN33} as magnetic pressure is too small. 

As discussed in last section, even magnetic buoyancy is not working in {\sf AGN33}, the radiative efficiency is comparable to the values in {\sf AGNB25}. This is because when accretion is driven by the density waves, the Reynolds stress decreases with height much slower compared with the density profile (Figure \ref{fig:vertical}). This means a significant fraction of dissipation happens in the low density regions, where the photon diffusion time is small. As long as the funnel region is not very optically thick, these photons can escape with the outflow. While in run {\sf AGN150} where the funnel becomes optically thick and all the dissipation is still very far away from the outflow region, radiative efficiency drops significantly.

\section{Discussion}
\label{sec:discussion}

\subsection{Excitation Mechanism of the Spiral Density Waves}

Based on our analysis in Section \ref{sec:stress}, the spiral density waves are clearly responsible for the large Reynolds stress we find from the simulations. However, it is still unclear physically what sets the pattern speed of the global density waves in the simulations. In section \ref{sec:spiral}, we have determined that the pattern speeds of the density waves in the four simulations correspond to the Keplerian rotation velocity at radii varying from $57$ to $100r_g$. We have also checked that there are no special asymmetrical structures at these radii that may be responsible for the excitation of density waves. The initial tori in these simulations have the same shape and are centered at the same radii for the three runs {\sf AGN150, AGN33} and {\sf AGNB25} but pattern speeds in the three runs correspond to the orbital rotation speed at quite different radii. The run {\sf AGNB52} has the initial torus closer to the central black hole but its pattern speed is smaller than the values in {\sf AGN33} and {\sf AGNB25}. These results are inconsistent with excitation of density waves by some special structure(s) in the initial torus region. Instead, it seems that the density waves are self-excited within the inner accretion flow itself.

Excitation of density waves is very common in accretion disks. In binary systems with the gravitational force from the companion star providing the perturbing force in the disks, the pattern speed of density waves can be easily associated with the orbital speed of the companion \citep[][]{ShuLubow1981,PapaloizouLin1995,Juetal2016}. Spiral density waves can also be excited without a perturbation at a fixed location, although density waves in this case may not be very coherent and less prominent compared with the previous case. For example, density waves are commonly observed in disks with self-gravitational instability \citep[][]{Dongetal2015,KratterLodato2016}, which also do not have a particular perturber at a fixed radius that may correspond to the pattern speed of the density waves. Turbulence from MRI can also excite spiral density waves \citep[][]{HeinemannPapaloizou2009a,HeinemannPapaloizou2009b}, which may explain how the density waves are excited in our simulations. However, it is unclear how the density waves excited at different radii form a coherent structure globally, which strongly motivates further study of this issue in future work.

It is well known that the propagation of density waves depends on the Mach number of the disk \citep[][]{Savonijeetal1994,Juetal2016,Juetal2017}, which is related to the ratio between the disk scale height and radius. In the super-Eddington accretion disks, the disk is thicker and density waves have longer wavelength. These spiral density waves propagate further and more efficiently transport angular momentum. For AGN disks with sub-Eddington accretion rates, the disk is thinner and Mach number will increase. Even if density waves are also excited there, they may have difficulty propagating. The role of density waves in these lower accretion rate disks will be a topic of future study.

\subsection{Angular Momentum Transfer with the Radiation Field}
\label{sec:radvis}
In the optically thick regime, radiating fluid in the co-moving frame can be described as a viscous 
gas with radiation providing the shearing viscosity \citep{MihalasMihalas1984,KaufmanBlaes2016} as (in the spherical polar coordinate)
\begin{eqnarray}
P_{r,vis}^{r,\phi}=\frac{8}{27}\frac{E_{r,0}}{\rho c^2}\frac{c}{\kappa_t}\left[
\frac{1}{r\sin\theta}\frac{\partial v_r}{\partial \phi}+r\frac{\partial }{\partial r}\left(\frac{v_{\phi}}{r}\right)
\right],
\end{eqnarray}
where $E_{r,0}$ is the radiation energy density in the co-moving frame and $\kappa_t\equiv \kappa_s+\kappa_{aR}$ is the total Rosseland mean opacity. Compared with radiation pressure, radiation viscosity is the order of $\mathcal{O}(\lambda_p v_{\phi}/rc)$ with $\lambda_p$ to be the photon mean free path across radius $r$. Radiation viscosity depends on the velocity gradient and can in principle transfer angular momentum outward along with the Maxwell stress and Reynolds stress. However, this is different from the off-diagonal component of the lab frame radiation pressure $P_{r}^{r\phi}$ 
in the total angular momentum flux equation \ref{eqn:ang}, which is related to the co-moving frame radiation pressure ${\sf P}_{r,0}$ as given by equation 91.12 of \cite{MihalasMihalas1984}. Even with isotropic radiation field in the co-moving frame, $P_{r}^{r\phi}$ still has a component proportional to the velocity as (to the order of $\mathcal{O}\left(v/c\right)^2$)
\begin{eqnarray}
P_r^{r,\phi}\approx \frac{4}{3}\frac{v_rv_{\phi}}{c^2}E_{r,0},
\label{eqn:prrphi}
\end{eqnarray}
which is non-zero for the disk with inflow $v_r$ and rotation $v_{\phi}$. Physically, this term multiplied by $r\sin\theta$ represents the angular momentum flux in the radiation field carried with the accretion flow, which is smaller than the hydrodynamic component $\rho v_rv_{\phi}$ by 
a factor of $E_{r,0}/\rho c^2$. When we estimate the \emph{outward} angular momentum flux caused by the radiation field in Section \ref{sec:radial}, we need to subtract the \emph{inward} component associated with $\<\rho v_r\>$ and $\< v_{\phi}\>$. This is similar to what we have done to determine the contribution of outward angular momentum transport due to the Reynolds stress. 

Because of the large optical depth in the super-Eddington accretion disks, $P_r^{r,\phi}$ is dominated by the component associated with the inflow gas, while radiation viscosity is much smaller than the Maxwell stress and Reynolds stress as shown in Figure \ref{fig:radial}. When accretion rate drops to the sub-Eddington regime the optical depth is expected to drop, increasing the photon mean free path.  At
the same time, the radiation pressure to gas pressure ratio will decrease.  The former increases
the importance of radiation viscosity while the latter decreases it.  If the increased mean free path
is more important, than radiation viscosity may play a larger role in angular momentum transport in the sub-Eddington flows, a topic we will examin in future study.

\subsection{Comparison with super-Eddington Accretion Disks  around Stellar Mass Black Holes}
The simulations presented here represent the first radiation magneto-hydrodynamics study of super-Eddington accretion onto supermassive black holes with realistic parameters. Many properties of these disks are unique outcome of the extremely large ratio between radiation pressure and gas pressure, which cannot happen for simulations designed for stellar mass black holes.

The significant Reynolds stress (comparable or even larger than the Maxwell stress) caused by the spiral shocks found in these AGN simulations is not observed for the stellar mass black hole case \cite{Jiangetal2014c}. This is also true for simulations designed for stellar mass black holes done by many other groups \citep[][]{Ohsugaetal2005,McKinneyetal2014,Sadowskietal2014}. The large radiation pressure is likely responsible for the suppression of Maxwell stress 
as well as the enhanced Reynolds stress because of the increased compressibility of the flow in this regime. For a similar net mass accretion rate in Eddington unit, the AGN disks have much larger turbulent inflow and outflow mass fluxes compared with the disks around stellar mass black holes. This can be confirmed by comparing the run {\sf AGNB52} at the bottom panel of Figure \ref{avemdot} and Figure 4 of \cite{Jiangetal2014c}. The mass flux as well as the kinetic energy flux associated with the unbound outflow in the AGN disks are also larger than the values in disks around stellar mass black holes. 

Despite the above differences, many properties of the super-Eddington accretion disks are common for both AGNs and stellar mass black holes. They both show strong radiation driven outflow along the funnel region with similar escaping velocity $\sim 0.3c-0.4c$ for disks with similar net mass accretion rate in Eddington unit. When the accretion rates are $\sim 25-50\Medd$, these simulations do confirm that the radiative efficiency can reach $\sim 4\%-6\%$ as in \cite{Jiangetal2014c}, although it is still unclear why it is different from the results found by others  \citep{McKinneyetal2014,Sadowskietal2014} in the similar accretion rate range.  In the case when MRI turbulence is the dominant mechanism for angular momentum transfer, we also see the butterfly diagram as well as the vertical advective cooling due to magnetic buoyancy as in \cite{Jiangetal2014c}. When accretion rate reaches $150\Medd$, the radiative efficiency drops significantly as suggested in the slim disk model. The differences are clearly shown in Figure \ref{averhov}. In this case, most of the funnel region becomes optically thick. The disk has inflow instead of outflow inside $\sim 50-60r_g$. As most of the dissipation still happens in this region, the photons are trapped and advected towards the black hole with a small fraction radiated due to the diffusive flux. 

\subsection{Observational Implications for Super-Eddington Accretion disk in AGNs}
\label{sec:obs}
Mid-plane temperatures of these super-Eddington AGN disks vary from $\sim 4\times10^5K$ to $8\times 10^5K$, which are much smaller than the 
temperature of disks around stellar mass black holes. This is consistent with the standard disk models \citep{ShakuraSunyaev1973,Abramowiczetal1980}. However, the optically thick outflow in the super-Eddington regime pushes the photosphere away from the disks, which reduces the effective temperature. The super-Eddington luminosity is achieved with larger emission area and smaller effective temperature compared with the standard disk models, which will likely change the observational appearance of these disks. Although we do not have theoretically calculated spectra for these disks yet, the effective temperature of the radiation field calculated by the simulations can give us an order of magnitude estimate of the expected spectrum peaks. For the runs {\sf AGN150, AGN33, AGNB25, AGNB52}, at the fixed height $z=1600r_g$, the radiation temperatures $T_r$ are $\sim 9.6\times 10^4K$, $5.0\times 10^4K$, $8.6\times 10^4K$ and 
$1.0\times 10^5K$ respectively inside $100r_g$ from the rotation axis. Because density near the outer boundary of the simulation domain for the run {\sf AGN150} is still much larger than the density of the other runs, the effective temperature at the true photosphere will likely be even smaller. %{\bf How this compared with NLS1 spectrum peaks?}

\begin{figure*}[htp]
	\centering
	\includegraphics[width=1.0\hsize]{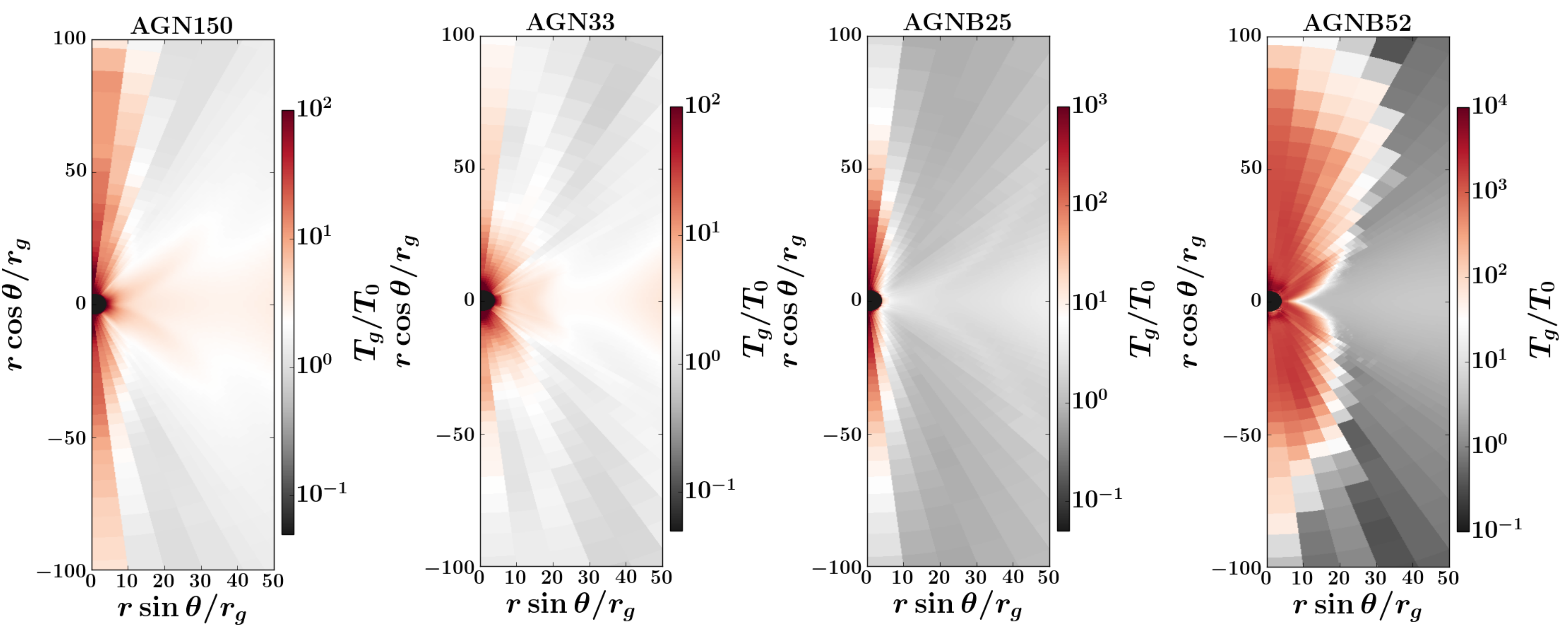}
	\caption{Time and azimuthally averaged spatial structures of gas temperature 
		$T_g$ in unit of $T_0=2\times 10^5K$. 
		From left to right, they are for runs {\sf AGN150, AGN33, AGNB25, AGNB52} 
		as labeled in each panel. The disks driven by spiral shocks do not have hot gas with 
  	   temperature significant above $10^7K$  as in {\sf AGN150} and {\sf AGN33}, while 
     {\sf AGNB52} shows hot gas with temperature up to $\sim 10^9K$.}
	\label{corona}
\end{figure*}

The above estimated effective radiation temperature cannot tell us the properties of X-rays from these disks, as 
X-rays from AGNs are generally believed to be produced by the Compton scattering with hot electrons. As shown in Figure \ref{fig:vertical}, 
thermal properties of the gas in the funnel regions are quite different for disks primarily driven by spiral shocks or MRI turbulence. 
This can also be seen clearly in Figure \ref{corona}, which shows the azimuthally averaged spatial distribution of gas temperature in the four runs. 
For the runs {\sf AGNM1} and {\sf AGNM2} with significant spiral density waves and very weak MRI turbulence, the gas is barely heated above 
$\sim 10^7K$.
% Therefore these disks will not have strong corona as well as X-ray photons. 
 The gas starts to be heated above  $10^8K$ in the run {\sf AGNB25} 
with a weak poloidal magnetic field and stronger MRI turbulence. Significant hot gas with temperature up to $\sim 10^9K$ shows up in the run {\sf AGNB52} 
with MRI turbulence to be the dominant mechanism for angular momentum transfer. However, despite the differences of gas temperature in the inner region of the disk, they are all embedded within the optically thick outflow in these super-Eddington disks. The total optical depths from the hot gas to the observer with most viewing angles are still much larger than $1$.  Therefore, we expect X-rays from super-Eddington AGN disks are generally much weaker. 

%X-ray photons are expected from this disk. Therefore, the amount of X-rays from AGN disks can be a way to distinguish the two different %mechanisms of accretion. 

Our super-Eddington accretion disks with accretion rate $\sim 20-50\Medd$ can have total radiative luminosity $\sim 10-20\Ledd$, which is larger than the 
luminosity predicted by the slim disk model. However, because of the drop of radiative efficiency with accretion rate $\sim 150\Medd$, the radiative luminosity from super-Eddington disks does not increase with increasing mass accretion rate indefinitely. In fact, luminosity from the run {\sf AGN150} is smaller than 
$10\Ledd$ despite the largest mass accretion rate in the four runs. This implies that after including the advective cooling along the vertical direction and 
realistic spatial distribution of dissipation, there will still be a maximum luminosity that the super-Eddington accretion disks can have, which is much larger than $\Ledd$ though.

%The outflows in these simulations have large density fluctuations, which can reduce the effective radiation acceleration as shown in Figure %\ref{fig:force}. 
%Recent observations of AGN accretion disks based on either micro-lensing \citep[][]{Morganetal2010,Chartasetal2016} or reverberation mapping %techniques \citep[][]{Edelsonetal2015,Jiangetal2017} infer that half light radius of the disk is larger than the corresponding value predicted by the %standard thin disk model. 
%\cite{DexterAgol2011} suggested that large azimuthal temperature fluctuations at each radius could explain the differences. It was unclear how %this large azimuthal temperature fluctuations can be maintained in the optically thick disk with rapid photon diffusion. The spiral shocks can be a %physical way to achieve this. This mechanism will also produce 
%correlated fluctuations at different radii because of the coherent structures of the spiral pattern. However, the simulations shown in this paper %have much larger accretion rates than the normally observed quasar accretion disks. In the super-Eddington regime, the spiral shocks are deeply %inside the photosphere. The large azimuthal temperature fluctuations get significantly damped when the photons reach the photosphere. It still %remains to be seen how important the spiral shocks are in AGN disks with sub-Eddington accretion rate as well as their effects on the %observational appearance. 

\section{Summary}
\label{sec:summary}
In summary, we have shown four 3D global radiation magneto-hydrodynamic simulations of super-Eddington accretion disks onto a $5\times 10^8\msun$ black hole for the first time. The accretion rates reach $\sim 25$ to $150\Medd$, leading to a radiation pressure to gas pressure ratio of $\sim 10^4-10^5$, which is expected only in AGN disks. The most important results we find from the simulations are summarized below.
\begin{itemize}
	\item When the accretion rate is $\sim 25-50\Medd$, outflow is launched along the funnel region starting from $5-10r_g$ with terminal velocity $\sim 0.3-0.4c$. In this case, the radiative efficiency can reach $\sim 5-7\%$. The dissipated energy in the main body of the disk can be efficiently advected to the funnel region and carried out by the outflow. The total kinetic energy luminosity associated with the funnel outflow is $\sim 15-30\%$ of the radiative luminosity. Mass flux carried by the outflow varies from $15\%-50\%$ of the net mass accretion rate in the disk.
	
	\item When the accretion rate reaches $\sim 150\Medd$, the optically thin funnel region with respect to the rotation axis is smaller than $5^{\circ}$. Outflow is launched from $\sim 50-70r_g$ with final speed $\sim 0.1-0.15c$. The radiative efficiency drops below $1\%$ with luminosity smaller than $15.5\Ledd$, because most of the energy is advected towards the black hole in this case. The  total kinetic energy luminosity carried with the outflow is $3.6\Ledd$ while 
	the outflow mass flux is $\sim 15\%$ of the net mass accretion rate. 
	
	\item Density waves are excited in the disks when gas flows towards the black hole from the initial torus. These density waves transfer angular momentum via spiral shocks. In the case when there are either no or only small amount of net poloidal magnetic fields, Maxwell stress produced by the MRI turbulence is suppressed in the strongly compressible radiation pressure dominated flow. Reynolds stress generated by the spiral shocks is the dominant mechanism for angular momentum transfer. Maxwell stress only becomes larger than the Reynolds stress when the ratio between radiation pressure and magnetic pressure associated with the mean poloidal magnetic fields $\<B_{\theta}\>^2/2$ is smaller than $\sim 10^4$.
	
	\item Although radiation pressure is $\sim 10^4-10^6$ times larger than the gas pressure and the radiation sound speed reaches $\sim 10\%$ speed of light, radiation viscosity plays a very minor role compared with Maxwell and Reynolds stresses for angular momentum transfer in these super-Eddington accretion disks because the large optical depth corresponds to a small mean free path.
	
	\item Super-Eddington accretion disks driven by spiral shocks do not produce hot gas in the funnel region. The gas temperature is always smaller than $\sim 10^7K$. For the disk with MRI turbulence being the dominant mechanism for angular momentum transfer, hot gas with temperature up to $10^9 K$ is observed. However, all the hot gas is covered by the optically thick outflow and super-Eddington AGN disks are expected to be X-ray weak. 
	
\end{itemize}

There are a few caveats of the simulations that we will improve in the future. We adopt the pseudo-Newtonian potential to mimic the general relativity effects around a non-spinning black hole, which means we cannot follow the formation of relativistic jets by spinning black holes. 
Extending our radiative transfer scheme to work in general relativity will be our next step. Adding frequency-dependent transport to the existing simulations will also be important to handle the Compton scattering self-consistently and treat the frequency dependence of the opacity accurately for AGN disks. The line-driven wind, which is believed to be important to drive the large scale outflows in most AGNs, is also not included in these simulations. It will be interesting to see how important it is for these super-Eddington disks. Lastly, we have started to post-process our simulations to compute synthetic spectra, which will be shown in a separate publication.

\section*{Acknowledgements}
We thank Omer Blaes, Geoffroy Lesur, Jeremy Goodman, Charles 
Gammie, Julian Krolik as well as many participants of the accretion disk 
program in KITP  for  helpful discussions. 
This research was supported in part by the National Science Foundation under Grant No. NSF PHY-1125915 
and AST-1333091. 
S.W.D. is supported by a Sloan Foundation Research Fellowship and a Virginia Space Grant Consortium New Investigator award.
An award of computer time was provided by the Innovative and Novel Computational Impact on Theory and Experiment (INCITE) program. 
This research used resources of the Argonne Leadership Computing Facility, 
which is a DOE Office of Science User Facility supported under Contract DE-AC02-06CH11357.
Resources supporting this work were also provided by the NASA High-End Computing (HEC) 
Program through the NASA Advanced Supercomputing (NAS) Division at Ames Research Center and
the Extreme Science and Engineering Discovery Environment (XSEDE), which is supported by
National Science Foundation (NSF) grant No. ACI-1053575.

%\clearpage

%\end{thebibliography}

\bibliographystyle{astroads}
\bibliography{SuperEddAGN}

\end{CJK*}

\end{document}